\documentclass[%
aps,
prd,
twocolumn,
preprintnumbers,
nofootinbib,
amsmath,amssymb,
]{revtex4-2}

\usepackage{style}
\usepackage{aas_macros}

\begin{document}

\preprint{DESY-25-134}

\title{Probing Quadratically Coupled\\ Ultralight Dark Matter with Pulsar Timing Arrays}

\author{Xucheng Gan\,\orcidlink{0000-0003-2834-7498}}
% \email{xucheng.gan@desy.de}
\affiliation{Deutsches Elektronen-Synchrotron DESY, Notkestr. 85, 22607 Hamburg, Germany}

\author{Hyungjin Kim\,\orcidlink{0000-0002-8843-7690}}
% \email{hyungjin.kim@desy.de}
\affiliation{Deutsches Elektronen-Synchrotron DESY, Notkestr. 85, 22607 Hamburg, Germany}

\author{Andrea Mitridate\,\orcidlink{0000-0003-2898-5844}}
% \email{andrea.mitridate@desy.de}
\affiliation{Deutsches Elektronen-Synchrotron DESY, Notkestr. 85, 22607 Hamburg, Germany}

\begin{abstract}
Ultralight dark matter may couple quadratically to Standard Model particles. Such quadratic interactions give rise to both coherent and stochastic signals in pulsar timing array (PTA) observations. In this work, we characterize these signals, including the effects of dark matter propagation in a finite-density medium, and assess the sensitivity of current and upcoming PTA observations to their detection. For coherent signals, we find that the sensitivity of current PTA observations competes with and sometimes exceeds that of other probes, such as equivalence principle tests and atomic clocks. For stochastic signals, we find that PTA sensitivities underperform equivalence principle constraints for both existing and upcoming PTA data sets.
\end{abstract}

\maketitle
\tableofcontents

\section{Introduction}
Pulsar timing arrays (PTAs) have been identified as a powerful probe of dark matter (DM), especially for DM particles with a mass $m_\phi\ll 1\,{\rm eV}$, known as ultralight dark matter (ULDM) candidates. Such dark matter candidates arise from various beyond the Standard Model scenarios, for instance, QCD axion models~\cite{Peccei:1977hh, Peccei:1977ur, Weinberg:1977ma, Wilczek:1977pj, Kim:1979if, Shifman:1979if, Zhitnitsky:1980tq, Dine:1981rt, Preskill:1982cy, Abbott:1982af, Dine:1982ah} and dynamical models for the electroweak hierarchy problem~\cite{Graham:2015cka, Arvanitaki:2016xds, Geller:2018xvz, Banerjee:2018xmn, Arkani-Hamed:2020yna, Banerjee:2020kww, TitoDAgnolo:2021nhd, TitoDAgnolo:2021pjo, Chatrchyan:2022pcb, Chatrchyan:2022dpy, Csaki:2024ywk}.

In this mass range, dark matter behaves as a classical field, oscillating at the frequency set by its mass. 
Oscillations in the field produce metric fluctuations that can be searched for using PTA observations. As metric perturbations are quadratically proportional to the underlying dark matter field, their oscillations exhibit two characteristic modes: \emph{fast modes}, associated with the coherent oscillation of the dark matter field, which induce metric perturbations oscillating in time with an angular frequency $\omega=2 m_\phi$~\cite{Khmelnitsky:2013lxt}. PTA searches for these coherent oscillations have placed constraints on the galactic abundance of dark matter with masses in the range $10^{-24}\,{\rm eV}- 10^{-22}\,{\rm eV}$~\cite{Porayko:2018sfa, NANOGrav:2023hvm, EPTA:2023xiy}. ULDM density fluctuations also exhibits \emph{slow modes}~\cite{Kim:2023pkx, Kim:2023pvt, Kim:2023kyy, Kim:2024xcr}, low-frequency stochastic fluctuations, related to the interference pattern of the ULDM field~\cite{Schive:2014dra, Hui:2021tkt}.\footnote{Throughout this work, we use fast-mode fluctuations and coherent signals interchangeably, and likewise slow-mode fluctuations and stochastic signals.} These slow modes induce stochastic metric fluctuations with a characteristic angular frequency of $\omega\lesssim m_\phi\sigma^2$, where $\sigma\simeq160\;{\rm km}/{\rm s}$ is the velocity dispersion of the virialized dark matter halo. Searches for the ULDM-induced stochastic PTA signal have placed constraints on the solar-system abundance of dark matter with masses $10^{-18}\,{\rm eV}-10^{-16}\,{\rm eV}$~\cite{Kim:2023kyy}.

Ultralight dark matter might also couple to the Standard Model non-gravitationally. When these couplings are linear in the dark matter field, ULDM coherent oscillations induce oscillations of fundamental constants of nature (see Refs.~\cite{Antypas:2022asj, Uzan:2024ded} for recent reviews). Such coherent oscillations affect the time and frequency standards that are used to record pulsar timing data~\cite{Graham:2015ifn, Kaplan:2022lmz}, and the rotational period of pulsars~\cite{Kaplan:2022lmz}, leaving coherent signals in the timing data. Searches for these coherent signals have placed constraints on the strength of linear couplings between Standard Model and ULDM field in the mass range $10^{-24}\,{\rm eV} - 10^{-22}\,{\rm eV}$~\cite{NANOGrav:2023hvm, EPTA:2023xiy, Kaplan:2022lmz}.

Non-gravitational interactions between ultralight dark matter and the Standard Model could be quadratic in the underlying dark matter field. Such quadratic interactions could arise, for example, in models of QCD axions and axion-like particles~\cite{Kim:2022ype,Beadle:2023flm,Kim:2023pvt,Gan:2023wnp,Gan:2025nlu}. In this case, timing signals are expected to exhibit both coherent and stochastic fluctuations, similarly to gravitational interactions. While both coherent and stochastic fluctuations from gravitational interactions have been extensively studied in the context of PTA observations, an analysis of quadratic non-gravitational interactions was only performed in Ref.~\cite{Smarra:2024kvv}, where the authors searched for the coherent signal produced by ULDM with a universal conformal coupling to gravity. This work generalizes these results by studying both the coherent and stochastic signals produced by an ULDM field with arbitrary quadratic couplings to the SM, and including the impact of matter effects on the PTA signals. Note that when dark matter form bound objects, quadratic interactions could give rise to transient signals in timing observation~\cite{Stadnik:2014cea}.

The paper is structured as follows. In Sec.~\ref{sec:uldm_signal}, we derive the PTA signals induced by the acceleration of Solar System objects and pulsars, by atomic clocks used to record times of arrival, and by fluctuations in pulsar spin rate, all due to the dark matter field surrounding the Solar System and pulsar. In Sec.~\ref{sec:pta_analysis}, we discuss how we implemented the Bayesian search for the signals discussed in Sec.~\ref{sec:uldm_signal}, and how we generated the synthetic data sets used in this work. The results of this Bayesian search are presented in Sec.~\ref{sec:results}, together with a discussion of the validity of some of the assumptions made in deriving them in Sec.~\ref{subsec:gaussianity}. We conclude in Sec.~\ref{sec:conclusions}. In App.~\ref{sec:gaussianity_appx}, we discuss the non-Gaussianity of the stochastic fluctuations. In App.~\ref{sec:matter_effect_appx}, we discuss the impact of the matter effect on the propagation of quadratically coupled dark matter around astrophysical objects and provide an updated limit on the light QCD axion from equivalence principle tests.

\section{Quadratic Interactions}
\label{sec:uldm_signal}

We consider a scalar field quadratically coupled to the Standard Model. The action is given by
\begin{equation}
    S = \int d^4x \, \left[
    \frac{1}{2} ( \partial_\mu \phi )^2
    - \frac{1}{2} \mphi^2 \phi^2 
    + {\cal L}_2
    \right],
\end{equation}
where $\mphi$ is the dark matter mass, and ${\cal L}_2$ is the Lagrangian describing the quadratic interactions between the ULDM field and the Standard Model. Following the conventions of Refs.~\cite{Damour:2010rm, Damour:2010rp, Hees:2018fpg, Banerjee:2022sqg}, we write this Lagrangian as:
\begin{align}\label{eq:lagrangian}
    {\cal L}_2
    = 
    \frac{\phi^2}{2\Mpl^2}
    \bigg[
    \frac{d_\gamma}{4e^2}F_{\mu\nu}F^{\mu\nu} 
    - \frac{d_g\beta_3}{2 g_3} G_{\mu\nu}^AG_A^{\mu\nu}
    \nonumber\\
    - \sum_\psi (d_{m_\psi}+\gamma_{m_\psi} d_g) m_\psi \bar \psi \psi
    \bigg]\,,
\end{align}
where $\beta_3$ is the beta function coefficient of the strong sector, $\gamma_{m_\psi}$ are the anomalous dimensions of the light quarks, $\Mpl=(4\pi G_N)^{-1/2}$ is the Planck scale, and $(d_g, d_\gamma,  d_{m_\psi})$ are dimensionless coupling constants. The sum over $\psi$ runs over the electron, and the up and down quarks. Due to these couplings, the values of the fundamental constants depend on the background ULDM field. For example, the fractional variations of the QCD scale, $\lambdaqcd$, the fine structure constant, $\alpha$, and the fermion masses, $m_\psi$, are given by
\begin{equation}\label{eq:lep_mas_shift}
    \frac{\delta \lambdaqcd}{\lambdaqcd} = d_g \varphi^2,
    \quad
    \frac{\delta\alpha}{\alpha}=d_\gamma \varphi^2,
    \quad
    \frac{\delta m_\psi}{m_\psi}=d_{m_\psi}\varphi^2\,,
\end{equation}
where we have defined the dimensionless quadratic scalar operator
\begin{equation}
    \label{eq:def_varphi_sq}
    \varphi^2 \equiv \frac{\phi^2}{2M_{\rm pl}^2}.
\end{equation}
Therefore, fluctuations in the ULDM field induce fluctuations in the values of fundamental constants.

The dark matter signals in timing observations will depend similarly on the quadratic field operator, and hence it is important to first characterize the statistical properties of the quadratic operator. The fluctuations in the quadratic operator can be described by its power spectrum,
\begin{equation}\label{eq:phi_ps}
    \langle\widetilde{\varphi^2}(k)\, \widetilde{\varphi^2}^*(k') \rangle = (2\pi)^4 \delta^{(4)}(k - k') P_{\varphi^2}(k) , 
\end{equation}
where we have introduced the Fourier component of the dimensionless quadratic ULDM operator, $\widetilde{\varphi^2}(f,\bm{k})$. The power spectrum consists of two parts~\cite{Kim:2023pvt}:
\begin{equation}\label{eq:phi_ps_decomp}
    P_{\varphi^2}(k) = P_{\varphi^2}^{\rm fast}(k) + P_{\varphi^2}^{\rm slow}(k)\,,
\end{equation}
where the first term arises from the coherent fluctuation at $\omega = 2 m_\phi$, and the second arises from the stochastic fluctuation at $\omega \lesssim m_\phi \sigma^2$. Note that the quadratic operator contains a zero mode, which contributes to an unobservable time-independent signal. In Eq.~\eqref{eq:phi_ps} and what follows, we denote $\varphi^2$ as the quadratic operator with its mean value subtracted, i.e., $\varphi^2 \to \varphi^2 - \langle \varphi^2\rangle$.

Under the assumption that dark matter velocity, $\bm{v}$, has an isotropic distribution,\footnote{Here we neglect the anisotropies in the DM velocity distribution induced by the Solar System's revolution around the Galactic Center. These anisotropies alter the overlap reduction function for the ULDM signal. However, as discussed in Ref.~\cite{Kim:2023kyy}, while they are important for model discrimination, they do not significantly alter the upper limits we can place on the ULDM signal.}
\begin{equation}
    f(\boldsymbol v)
    = \frac{\bar \rho / \mphi}{(2 \pi \sigma^2)^{3/2}}
    \exp\left( - \frac{|\bm{v}|^2}{2\sigma^2} \right)\,,
\end{equation}
the power spectrum can be analytically computed as
\begin{align}
    P_{\varphi^2}^{\rm fast}(k) &= 
    \frac{1}{4\Mpl^4}
    \frac{2\pi^2\bar\rho^2}{m_\phi^8\sigma^5}
    e^{-\frac{\omega-2m_\phi}{m_\phi\sigma^2}}
    \sqrt{\frac{\omega-2m_\phi}{m_\phi\sigma^2}-\frac{|\bm{k}|^2}{4 m_\phi^2 \sigma^2}}
    ,
    \\
    P_{\varphi^2}^{\rm slow}(k) &=
    \frac{1}{4\Mpl^4} \frac{2\pi^2 \bar\rho^2}{|\bm{k}| m_\phi^7 \sigma^4}
    \exp\bigg[
    -\frac{|\bm{k}|^2}{4m_\phi^2\sigma^2} - \frac{\omega^2}{\sigma^2|\bm{k}|^2}
    \bigg]
    .
\end{align}
where $k\equiv(\omega, \bm{k})$ denotes the scalar four-momentum, $\bar\rho \simeq 0.4\,{\rm GeV}/{\rm cm}^3$ denotes the mean dark matter density, and $\sigma$ denotes the velocity dispersion. The power spectra in this work are one-sided, defined only for positive frequencies, $\omega \geq 0$. The goal of the remainder of the section is to relate the quadratic spectrum above to that of the ULDM-induced timing signal. Specifically, we are interested in three potential signals:

\bigskip
\noindent\textbullet~\emph{Doppler signal:} The ULDM-induced perturbations to fundamental constants cause the masses of pulsars and Solar System objects to depend on the external ULDM field. Because the dark matter has a non-vanishing spatial gradient, an object of mass $M[\varphi^2]$ experiences an additional force given by
\begin{equation}
    \pmb{\cal F}(t) = - \frac{\vecnabla M}{M} \propto \vecnabla\varphi^2 \,,
    \label{force_eq}
\end{equation}
where $\pmb{\cal F}$ is force per unit mass. This additional force introduces a Doppler redshift in the perceived rotational period of the $a$-th pulsar given by
\begin{equation}
    z_a(t) \equiv 
    \left( \frac{\delta P}{P} \right)_a
    = \nhat_a \cdot 
    \big[
    \delta \bm{v}_{a}(t- L_a) - \delta \bm{v}_{\odot}(t)
    \big]\,,
    \label{redshift_doppler}
\end{equation}
where $P_a$ is the perceived period of the $a$-th pulsar, $\nhat_a$ is a unit vector pointing to the pulsar, and $\delta \bm{v}_{\odot}$ and $\delta \bm{v}_{a}$ are the velocity perturbations of the Solar System barycenter and pulsar $a$ respectively, which can be derived by solving Eq.~\eqref{force_eq}.

\medskip

\noindent\textbullet~\emph{Clock signal:} The pulsar period is measured with respect to Terrestrial Time~(TT), which is defined through an ensemble of atomic clocks. The reference frequency provided by Terrestrial Time, $f_{\rm TT}$, is determined by the transition frequencies of these clocks. Consequently, any perturbations to the underlying atomic transitions are directly imprinted on the measured pulsar period. Fluctuations in the fundamental constants—arising from the couplings in Eq.~\eqref{eq:lagrangian}—modify the atomic clock transition frequencies, thereby inducing fluctuations in $f_{\scriptscriptstyle\rm TT}$. These fluctuations manifest as a characteristic PTA redshift signal of the form:
\begin{equation}\label{eq:clock_signal}
    z_a(t) 
    = \frac{\delta f_{\scriptscriptstyle\rm TT}}{f_{\scriptscriptstyle\rm TT}}\propto \varphi^2\,,
\end{equation}
i.e., if the clock ticks faster, the perceived pulsar rotational period becomes longer, and vice versa.

\medskip

\noindent\textbullet~\emph{Pulsar spin signal:} Fluctuations in the pulsar constituent particle masses induce fluctuations in the pulsar moment of inertia, $I_a$. By the conservation of angular momenta, these fluctuations will induce the corresponding fluctuations in the pulsar spin frequency, $f_a$, leading to a PTA redshift signal of the form\footnote{Although we have separated the clock signal from the pulsar spin signal for convenience, these signals should be understood as a combined effect. Since times of arrival are always measured against Terrestrial Time, timing observations depend on the relative stability of two frequency standards, rather than on an individual one. Accordingly, the clock and pulsar spin signals could be combined into a single signal of the form $z_a(t) = \delta( f_{\scriptscriptstyle\rm TT} / f_a) / (f_{\scriptscriptstyle\rm TT}/f_a)$.}
\begin{equation}
    z_a(t) = \frac{\delta I_a}{I_a} =- \frac{\delta f_a}{f_a}\propto\varphi^2.
\end{equation}

\medskip

The timing residual, $\delta t_a$, induced by any of the effects mentioned above, can be expressed as:
\begin{equation}
    \delta t_a(t) = \int^t dt'  \, z_a(t'). 
\label{residual}
\end{equation}
Since all three signals discussed above are proportional to $\varphi^2$ or its gradient, their induced PTA signals will have the general form
\begin{equation}\label{eq:dt_general}
    \delta t_a(t) = \int \frac{d^4k}{(2\pi)^4}\;\mathcal{K}_a(f,\bm{k}) \;\widetilde{\varphi^2}(f,\bm{k}) e^{-2\pi i ft} , 
\end{equation}
where we have set the center of the coordinate system to be at the location of the observer, ${\bm{x}=0}$, without loss of generality. The kernel function, $\mathcal{K}_a(f,\bm{k})$, encodes the PTA response to each effect mentioned above. The specific form of the kernel functions will be derived in the rest of this section.

From Eq.~\eqref{eq:phi_ps} and Eq.~\eqref{eq:dt_general}, we find the two-point function for the ULDM-induced timing residuals as
\begin{equation}
    \langle \delta t_a(t)\delta t_b(t')\rangle = 
    \int_0^{\infty} df\,S^{\scriptscriptstyle\rm DM}_{ab}(f) \cos[2\pi f(t-t')]\,,
\end{equation}
where the power spectrum, $S^{\scriptscriptstyle\rm DM}_{ab}(f)$, is given by
\begin{equation}\label{eq:dt_ps_general}
    S^{\scriptscriptstyle\rm DM}_{ab}(f) = 
    2 \int \frac{d^3\bm{k}}{(2\pi)^3}\;\mathcal{K}_a(f,\bm{k})
    \mathcal{K}^*_b(f,\bm{k})
    P_{\varphi^2}(f,\bm{k})\,.
\end{equation}
This expression relates the spectrum of the quadratic operator $P_{\varphi^2}(k)$ and the kernel to the timing-residual spectrum due to each aforementioned effect. Once the kernel is identified, the computation of the spectrum becomes straightforward. 

In the following discussions, we show that Eq.~\eqref{eq:dt_ps_general} can be factorized as
\begin{equation}
    S^{\scriptscriptstyle\rm DM}_{ab}(f)=\Gamma_{ab}^{\scriptscriptstyle\rm DM} \, S^{\scriptscriptstyle\rm DM}(f)\,.
    \label{Spectrum_Angular}
\end{equation}
Here, $\Gamma_{ab}^{\scriptscriptstyle\rm DM}$ denotes the overlap reduction function, which encodes the angular correlations between pulsars $a$ and $b$, and $S^{\scriptscriptstyle\rm DM}(f)$ denotes the power spectral density that characterizes the frequency dependence of the signal. The power spectral density can be further decomposed into the contributions from the fast and slow modes:
\begin{equation}
    S^{\scriptscriptstyle\rm DM}(f)=S^{\scriptscriptstyle\rm DM}_{\rm fast}(f)+S^{\scriptscriptstyle\rm DM}_{\rm slow}(f)\,.
\end{equation}
We next derive explicit expressions for the fast- and slow-mode power spectral densities for the Doppler, clock, and pulsar spin signals.

\begin{table}[t]
    \bgroup
    \renewcommand{\arraystretch}{1.5}
    \setlength\tabcolsep{6pt}
    {\footnotesize
    \begin{tabular}{lccccc}
    \toprule
    \textbf{} 
    $\bm{d}$
    & $d_g$ 
    & $d_\gamma$ 
    & $d_{\hat m}-d_g$ 
    & $d_{\delta m}-d_g$ 
    & $d_{m_e}-d_g$ 
    \\[-4pt]
    \textbf{} 
    & 
    & [$\times 10^{-4}$]
    & [$\times 10^{-2}$]
    & [$\times 10^{-3}$]
    & [$\times 10^{-4}$]
    \\ \hline
    \multicolumn{6}{l}{\textbf{Particles}} \\[-2pt]
    $\bm{Q}_{n}$     & $1$ & $-1.4$ & $4.8$ & $1.7$  & $0$ \\
    $\bm{Q}_{p}$     & $1$ & $6.7$  & $4.8$ & $-1.7$ & $0$ \\
    $\bm{Q}_{e}$     & $1$ & $0$    &  $0$  &  $0$   & $1 \times 10^4$ \\
    $\bm{Q}_{\alpha}$& $1$ & $5.1$  & $7.0$ & $0$    & $0$ \\
    \specialrule{0.2pt}{1pt}{1pt}
    \multicolumn{6}{l}{\textbf{Astrophysical Objects}} \\[-2pt]
    $\bm{Q}_{\odot}$     & $1$ & $6.3$ & $5.4$ & $-1.2$ & $4.7$ \\
    $\bm{Q}_{\oplus}$    & $1$ & $19$ & $8.1$ & $3.9 \times 10^{-2}$ & 2.7 \\
    $\bm{Q}_{\rm psr}$   & $1$ & $-0.59$ & $4.8$ & $1.4$ & $0.54$ \\
    \specialrule{0.2pt}{1pt}{1pt}
    \multicolumn{6}{l}{\textbf{Terrestrial Time}} \\[-2pt]
    $\bm{Q}_{\scriptscriptstyle\rm TT}$ & $1$ & $4.8\times 10^4$ 
    & $-3.9$ 
    & $1.7$ 
    & $2 \times 10^4$ \\
    \specialrule{0.2pt}{1pt}{1pt}
    \multicolumn{6}{l}{\textbf{Pulsar Moment of Inertia}}  \\[-2pt]
    $\bm{Q}_I$           
    & $-5$
    & $7.3$
    & $-24$
    & $-8.6$
    & $0.18$
    \\
    \bottomrule
    \end{tabular}
    }
    \egroup
    \caption{Effective charge vectors for particles, astrophysical objects, the frequency standard (Terrestrial Time), and the pulsar moment of inertia.}
    \label{tab:charge_vectors}
\end{table}

\subsection{Doppler Signal}\label{subsec:doppler_signal}
The force per unit mass acting on an object of mass $M$ due to ULDM fluctuation can be parametrized as
\begin{equation}
    \pmb{\cal F} (t) = - \frac{\vecnabla M}{M} = - g \, \vecnabla \varphi^2
    \label{parametrization_doppler}
\end{equation}
where $g$ is the coupling strength of ULDM to the object $M$. More specifically, this coupling is defined as~\cite{Damour:2010rp}
\begin{equation}
    \label{eq:charge_mass_vary}
    g
    = \frac{\partial \ln M}{\partial \varphi^2}
    = \bm{d} \cdot \bm{Q}\,,
\end{equation}
where $\bm{Q}$ is the charge vector of the object, and $\boldsymbol d$ is the basis of the quadratic couplings introduced in Eq.~\eqref{eq:lagrangian}:
\begin{equation}
    \bm{d}
    =
    (
        d_g, \,
        d_\gamma, \,
        d_{\hat m} - d_g, \,
        d_{\delta m} - d_g, \,
        d_{m_e} - d_g
        )\,.
\end{equation}
Here we have introduced
\begin{align}
    d_{\hat m} &= 
    \frac{d_{m_d} m_d + d_{m_u} m_u}{m_u + m_d}\,,
    \\
    d_{\delta m} &= 
    \frac{d_{m_d} m_d - d_{m_u} m_u}{m_d - m_u}\,,
\end{align}
with $m_u$ and $m_d$ being the up- and down-quark masses, and $\hat{m} = (m_u + m_d)/2$ and $\delta m = m_d - m_u$ denoting their symmetric and antisymmetric combinations, respectively.

With the above parametrization, we find that the kernel for the Doppler signal can be expressed as
\begin{equation}
    \label{eq:ker_dop}
    \mathcal{K}_a(f,\bm{k})
    = - \frac{g_\odot}{(2\pi f)^2} 
    (i \boldsymbol k \cdot \nhat_a )\, U_a,
\end{equation}
where
\begin{equation}
    U_a = 1 - (g_a / g_\odot) e^{2\pi i f L_a} e^{i \boldsymbol k \cdot \nhat_a L_a}.
\end{equation}
The first term in the $U_a$ function is related to what is commonly referred to as the \emph{Earth term} in the PTA literature, and corresponds to the ULDM signal due to the perturbation of Solar System.\footnote{Although it is referred to as the Earth term, we use the coupling between dark matter and the Sun for the Doppler signal. This choice is justified because the Earth term originates from the perturbation of the Solar System barycenter, which is dominated by the displacement of the Sun. In contrast, the contribution from perturbations in the Earth's position is suppressed by a factor of $M_\oplus/M_\odot$.} The second term, known as the \emph{pulsar term}, corresponds to the ULDM signal imprinted at the pulsar location. Here, $g_\odot$ and $g_a$ denote the effective coupling strengths of ULDM to the Sun and pulsar $a$, respectively. These couplings are given by
\begin{equation}
    g_{\odot}
    = \boldsymbol d \cdot \bm{Q}_\odot\,, \, \quad g_{a}
    = \boldsymbol d \cdot \bm{Q}_\psr\,,
\end{equation}
where $\bm{Q}_\odot$ and $\bm{Q}_\psr$ are the charge vectors of sun and pulsar, respectively.

The charge of each astrophysical object is determined by its composition. In particular, the charge vector of an object $\mathcal{A}$ can be expressed as
\begin{equation}
    \bm{Q}_\mathcal{A} = \sum_i \mathcal{X}_{\mathcal{A},i} \bm{Q}_i
\end{equation}
where ${\mathcal X}_{\mathcal{A},i}$ and $\bm{Q}_i$ are the mass fraction and the charge of the particle species $i$, respectively. Here, $\mathcal{A}$ labels the astrophysical objects such as Sun, Earth, and pulsar. The Sun is primarily composed of protons, alpha particles, and electrons with an approximate mass fraction of $\mathcal{X}_{\odot,p} \sim 0.7$, $\mathcal{X}_{\odot,\alpha} \sim 0.3$, and $\mathcal{X}_{\odot,e} \sim 5 \times 10^{-4}$, respectively~\cite{Asplund:2009fu}. On the other hand, a typical pulsar consists of neutrons, protons, muons, and electrons with a mass fraction of $\mathcal{X}_{\psr,n} \sim  0.9$, $\mathcal{X}_{\psr,p} \sim  0.1$, $\mathcal{X}_{\psr,\mu} \sim  0.005$, and $\mathcal{X}_{\psr,e} \sim 5\times 10^{-5}$~\cite{Bell:2019pyc}. The charge vectors of these astrophysical objects, including that of the Earth, are summarized in Table~\ref{tab:charge_vectors}.

Using the kernel~\eqref{eq:ker_dop}, we find the spectrum for the Doppler signal,
\begin{align}
        S^{\scriptscriptstyle\rm DM}_{\rm fast}(f) 
    &= 
    \frac{\pi}{3} \frac{g_\odot^2\bar\rho^2}{M_{\rm pl}^4m_\phi^3\omega^4}\left(\frac{\omega-2m_\phi}{m_\phi\sigma^2}\right)^3e^{-\frac{\omega-2m_\phi}{m_\phi\sigma^2}}, \label{eq:dop_fast}
    \\
    S^{\scriptscriptstyle\rm DM}_{\rm slow}(f) 
    &=
    \frac{4}{3} 
    \frac{g_\odot^2\bar\rho^2}{M_{\rm pl}^4m_\phi^3\omega^4}
    \left(\frac{\omega}{m_\phi\sigma^2}\right)^2
    K_2\left( \frac{\omega}{m_\phi\sigma^2}\right)\,,\label{eq:dop_slow}
\end{align}
and the overlap reduction function, 
\begin{equation}\label{eq:Gamma_ab_doppler}
    \Gamma_{ab}^{\scriptscriptstyle\rm DM}= \frac{1}{2}
    \Big[ 
    (g_a / g_{\odot})^2 \delta_{ab} + \nhat_a \cdot \nhat_b \Big]\,.
\end{equation}
If $g_a = g_{\odot}$, the overlap reduction function in Eq.~\eqref{eq:Gamma_ab_doppler} reduces to one with a dipolar pattern, like the one observed for the purely gravitational ULDM stochastic signals~\cite{Kim:2023kyy}. The above result ignores the pulsar term in $U_a$; the pulsar terms only lead to uncorrelated signals for slow-mode fluctuations, but might provide correlated signals for fast-mode fluctuations~\cite{Boddy:2025oxn}.

\subsection{Clock Signal}\label{subsec:clock_signal}

The clock redshift signal is given by
\begin{equation}\label{eq:clock_shift}
    z(t) = \frac{\delta f_{\scriptscriptstyle\rm TT}}{f_{\scriptscriptstyle\rm TT}} = 
    g_{\scriptscriptstyle\rm TT} \, \varphi^2,
\end{equation}
where $f_{\scriptscriptstyle\rm TT}$ is the frequency standard used to measure the times of arrival of radio pulses, and $g_{\scriptscriptstyle\rm TT}$ is the effective coupling strength parameterizing the dependence of $f_{\scriptscriptstyle\rm TT}$ on the underlying ULDM fluctuations. 

The subscript ``${\scriptscriptstyle\rm TT}$'' refers to Terrestrial Time, which is used by PTA collaborations as the reference timescale for time-of-arrival (TOA) measurements. Since Terrestrial Time is constructed in a way that closely follows the definition of SI second, which is given by the ground-state hyperfine transition in a cesium atom, we assume its frequency is identical to that of the Cs frequency standard, i.e. $f_{\scriptscriptstyle\rm TT} = f_{\rm Cs}$. The hyperfine frequency of a specific atom is
\begin{equation}\label{eq:f_hyperfine}
    f_{\rm hfs} = \frac{g_N \, m_e^2 \, \alpha^{4 + K} }{m_p}\,, 
\end{equation}
where $g_N$ is a nuclear $g$-factor, $m_p$ is the proton mass, and $K$ is the relativistic correction. For $^{133}{\rm Cs}$, the relativistic correction is $K_{\rm Cs}=0.83$~\cite{Dzuba:1998au}, and the variation of $g$-factor is~\cite{Flambaum:2006ip}
\begin{equation}\label{eq:g_factor_variation}
    \frac{\delta g_N}{g_N} = 0.009 \, \frac{\delta (\hat m / \lambdaqcd)}{(\hat m / \lambdaqcd)} = 0.009 \, ( d_{\hat m} - d_g ) \varphi^2\,.
\end{equation}
Consequently, the effective coupling strength for the clock signal can be written as
\begin{equation}
\label{eq:g_tt_main}
    g_{\scriptscriptstyle\rm TT} = \bm{d} \cdot \bm{Q}_{\scriptscriptstyle\rm TT},    
\end{equation}
where $\bm{Q}_{\scriptscriptstyle\rm TT}$ is the corresponding charge vector. The numerical value of $\bm{Q}_{\scriptscriptstyle\rm TT}$ is given in Table~\ref{tab:charge_vectors}.

The kernel associated with the clock signal takes a simple form:
\begin{equation}
    \mathcal{K}_a(f,\bm{k}) = \frac{i g_{\scriptscriptstyle\rm TT}}{2 \pi f}\,.
\end{equation}
Since this kernel is independent of the pulsar index, the overlap reduction function for the clock signal takes a monopolar form:
\begin{equation}\label{eq:Gamma_ab_clock}
    \Gamma_{ab}^{\scriptscriptstyle\rm DM}=1.
\end{equation}
This is because all times of arrival are measured with a common frequency standard. The power spectral densities of the timing residuals for the fast and slow modes are then given by
\begin{align}
    S^{\scriptscriptstyle\rm DM}_{\rm fast}(f) 
    &= 
    \frac{\pi g_{\scriptscriptstyle\rm TT}^2\bar\rho^2}{4M_{\rm pl}^4} \frac{1}{m_\phi^5\omega^2\sigma^2}\left(\frac{\omega-2m_\phi}{m_\phi\sigma^2}\right)^2e^{-\frac{\omega-2m_\phi}{m_\phi\sigma^2}} , 
    \label{eq:clock_fast}
    \\
    S^{\scriptscriptstyle\rm DM}_{\rm slow}(f) 
    &=
    \frac{g_{\scriptscriptstyle\rm TT}^2\bar\rho^2}{M_{\rm pl}^4}
    \frac{1}{ m_\phi^5\omega^2\sigma^2}
    \left(\frac{\omega}{m_\phi\sigma^2}\right)
    K_2\left( \frac{\omega}{m_\phi\sigma^2}\right)\,.\label{eq:clock_slow}
\end{align}

Comparing \cref{eq:dop_fast,eq:dop_slow} with \cref{eq:clock_fast,eq:clock_slow}, and recalling that the signal power is concentrated mainly at $\omega = 2 \mphi$ for the fast mode and at $\omega \lesssim \mphi \sigma^2$ for the slow mode, we find that the relative size of the Doppler and clock signals is parametrically given by
\begin{equation}
    \frac{S^{\rm clock}(f)}{S^{\rm Doppler}(f)}\sim
    \bigg( \frac{g_{\scriptscriptstyle\rm TT}}{g_\odot} \bigg)^2
    \times
    \begin{cases}
        \sigma^{-2} & \textrm{fast mode}
        \\
        \sigma^2  & \textrm{slow mode}
    \end{cases}. 
\end{equation}
This suggests that the clock signal is dominant for the fast mode, while the Doppler signal is dominant for the slow mode. 

\subsection{Pulsar Spin Signal}\label{subsec:psr_spin_signal}
The redshift signal due to ULDM-induced fluctuations in the pulsar's moment of inertia is given by
\begin{equation}\label{eq:pulsar_redshift}
    z_a(t) =\frac{\delta I_a}{I_a}=
    g_I \, \varphi^2(t,\bm{x}_a)\,,
\end{equation}
where $I_a$ is the moment of inertia of pulsar $a$, and $g_I$ is the effective coupling characterizing its dependence on the ULDM field value. To derive the above equation, we use the conservation of the pulsar spin, $S = I \omega$, to express the redshift as a variation of the pulsar moment of inertia.

For simplicity, we model the pulsar as a uniform sphere, with $I = (2/5)MR^2$, where $M$ and $R$ denote its mass and radius. Therefore, the pulsar spin signal can be expressed as~\cite{Kaplan:2022lmz}
\begin{equation}\label{eq:z_M_R_mn}
    z_a(t) 
    = \frac{\delta M}{M} + 2 \frac{\delta R}{R}
    = \frac{1}{3} \frac{\delta M}{M} 
    - \frac{16}{3} \frac{\delta m_n}{m_n}\,.
\end{equation}
In the second equality of Eq.~\eqref{eq:z_M_R_mn}, we use $R \propto M^{-1/3} \, m_n^{-8/3}$, which is obtained by modeling a pulsar as an object whose structure is maintained by the balance between the degenerate pressure of non-relativistic neutrons and the gravitational attraction. From Eq.~\eqref{eq:z_M_R_mn}, the effective coupling can therefore be written as
\begin{equation}\label{eq:g_I_main}
    g_{I} = \bm{d} \cdot \bm{Q}_I,
\end{equation}
where the charge vector is $\bm{Q}_I = (1/3)\,\bm{Q}_\psr - (16/3)\,\bm{Q}_n$. Its numerical value is summarized in Table~\ref{tab:charge_vectors}.

As can be seen from Eq.~\eqref{eq:pulsar_redshift}, the signal induced by fluctuations in the pulsar's moment of inertia has the same dependence on the ULDM field as the clock signal in terms of the power spectral densities. Specifically, the timing-residual power spectra for the pulsar spin signal are the same as those in Eqs.~\eqref{eq:clock_fast} and \eqref{eq:clock_slow}, except for the substitution
\begin{equation}
    g_{\scriptscriptstyle\rm TT}\to g_I.
\end{equation}
However, unlike the clock signal,  $\Gamma_{ab}^{\scriptscriptstyle\rm DM}$, which reflects the correlation pattern of the spin signal, is not given by Eq.~\eqref{eq:Gamma_ab_clock}, but instead depends on the coherence length of the ULDM field, as we will discuss in the next section (see the discussion below Eq.~\eqref{eq:amps}). 

\section{Analysis}\label{sec:pta_analysis}
The statistical tools and the noise modeling used in this analysis closely follow those already described in Refs.~\cite{NANOGrav:2018hou, Kaplan:2022lmz, Kim:2023kyy}. In this section, we provide a short review of these methods and discuss how we used them to search for the dark matter signals discussed in the previous section.
 
\subsection{Noise Modeling and the PTA Likelihood}\label{subsec:pta_likelihood}
Pulsar timing residuals are typically modeled as a combination of deterministic signals and stochastic processes. Stochastic processes are assumed to be zero-mean Gaussian processes, such that they are fully characterized by the timing-residuals two-point function:
\begin{equation}
    \label{eq:timing_2p}
    \langle\bm{\delta t}_{a}\bm{\delta t}_{b}\rangle 
     = \delta_{ab}\bm{N}_a +\bm{F}_a\left(\delta_{ab}\bm{\varphi}_a+\Gamma_{ab}^{\scriptscriptstyle\rm GW}\bm{\phi}^{\scriptscriptstyle\rm GW}+\Gamma_{ab}^{\scriptscriptstyle\rm DM}\bm{\phi}^{\scriptscriptstyle\rm DM}\right)\bm{F_b}^T ,
\end{equation}
where $\bm{\delta t}_a$ is a vector of size $N_{{\rm TOA},a}$ containing the measured timing residuals for the $a$-th pulsar. We now detail each term below.

The first term on the right-hand side of Eq.~\eqref{eq:timing_2p} encodes the white noise contribution to the timing residuals in a $N_{{\rm TOA},a}\times N_{{\rm TOA},a}$ matrix, $\bm{N}_a$. This matrix is typically expressed in terms of EFAC, $G_\mu$, EQUAD, $Q_\mu$, and ECORR, $J_\mu$, parameters (one for each backend, here indicated by the $\mu$ index):
\begin{equation}
    N_{a,ij}=G_{a,\mu}^2(\sigma_{a,i}^2+Q_{a,\mu}^2)\delta_{ij}\,,
    \label{eq:wn_matrix}
\end{equation}
 where $\sigma_i$ is the TOA uncertainty for the $i$-th TOA belonging to the backend $\mu$.\footnote{For the analysis of the NANOGrav 12.5-year data, which was derived with narrowband observations, we also include correlated noise across radio frequencies as
\begin{equation}
    \bm{N}_a\supset J_{a,\mu} \delta_{e(i)e(j)}\,
\end{equation}
where $J_\mu$ is the ECORR parameter, and $\delta_{e(i)e(j)}$ denotes a Kronecker delta that equals 1 only when the epochs are the same for both TOAs considered and 0 otherwise}

\begin{table}[t]
\bgroup
\renewcommand{\arraystretch}{1.5}
\setlength\tabcolsep{3.pt}
{\footnotesize
	\begin{tabular}{ccc}
		\toprule
		\textbf{Parameter}	                       &   \textbf{Description}                 &   \textbf{Prior}   \\ \hline
		$\log_{10} A_a$                            &   intrinsic red-noise amplitude$^*$        &   $\mathcal{U}(-20,\,-11)$    \\
		$\gamma_a$                                 &   intrinsic red-noise spectral index$^*$   &   $\mathcal{U}(0,\,7)$        \\ \midrule
		$\log_{10} A_{\scriptscriptstyle \rm GW}$  &   GWB signal amplitude                 &   $\mathcal{U}(-20,\,-11)$    \\
		$\gamma_{\scriptscriptstyle \rm GW}$       &   GWB signal spectral index            &   $\mathcal{U}(0,\,7)$        \\ \midrule
		$\log_{10}(m_\phi/{\rm eV})$                    &   DM mass                              &   $\mathcal{U}(-19,\,-12)$    \\
		$\sigma\;[{\rm km/s}]$                                   &   DM velocity dispersion               &   $\mathcal{U}(150,\,250)$    \\ 
        \bottomrule
	\end{tabular}
}
\egroup
\caption{Prior distributions for the parameters used in the stochastic ULDM searches presented in this work. The parameters marked with a $^*$ indicate parameters that are pulsar-dependent. In the last column, $\mathcal{U}(a,b)$ indicates a uniform distribution between $a$ and $b$.}
\label{tab:priors_stoch}
\end{table} 
\begin{table}[t]
\bgroup
\renewcommand{\arraystretch}{1.5}
\setlength\tabcolsep{3.pt}
{\footnotesize
	\begin{tabular}{ccc}
		\toprule
		\textbf{Parameter}	                       &   \textbf{Description}                 &   \textbf{Prior}   \\ \hline
		$\log_{10} A_a$                            &   intrinsic red-noise amplitude$^*$        &   $\mathcal{U}(-20,\,-11)$    \\
		$\gamma_a$                                 &   intrinsic red-noise spectral index$^*$   &   $\mathcal{U}(0,\,7)$        \\ \midrule
		$\log_{10} A_{\scriptscriptstyle \rm GW}$  &   GWB signal amplitude                 &   $\mathcal{U}(-20,\,-11)$    \\
		$\gamma_{\scriptscriptstyle \rm GW}$       &   GWB signal spectral index            &   $\mathcal{U}(0,\,7)$        \\ \midrule
        $\log_{10}(\hat A_E/{\rm s})$                                   &   ULDM Earth signal amplitude               &   $\mathcal{U}(-9,\,-4)$    \\
        $\log_{10}(\hat A_{P}/{\rm s})$                                   &   ULDM pulsar signal amplitude               &   $\mathcal{U}(-9,\,-4)$    \\
        $\log_{10}(m_\phi/{\rm eV})$                    &   DM mass                              &   $\mathcal{U}(-24,\,-18)$    \\
        $\hat\phi$                    &   Earth normalized signal amplitude                              &   $e^{-x}$    \\
        $\gamma_E$                    &   Earth signal phase                              &   $\mathcal{U}(0,2\pi)$    \\
        $\gamma_{P,a}$                    &   pulsar signal phase$^*$                              &   $\mathcal{U}(0,2\pi)$    \\
        \bottomrule
	\end{tabular}
}
\egroup
\caption{Prior distributions for the parameters used in the searches for deterministic ULDM signals presented in this work. Here we have defined the rescaled ULDM signal amplitudes, $\hat A_{E,P}\equiv A_{E,P}/\hat\phi$. The parameters marked with a $^*$ indicate parameters that are pulsar-dependent. In the prior column, values between square brackets indicate the ranges of uniform distributions.}
\label{tab:priors_det}
\end{table} 

The bracketed terms on the right-hand side of Eq.~\eqref{eq:timing_2p} represent the contributions to the timing residuals from all stochastic processes with long-timescale correlations. In our case, these include pulsar-intrinsic noise, the gravitational wave background (GWB), and potential DM signals. Such time-correlated processes are decomposed into a discrete Fourier basis with frequencies $f_k \equiv k/T_{\textrm{obs}}$, where $k$ labels the harmonics and $T_{\textrm{obs}}$ is the total observation span. The components of this basis are collected in the $N_{{\rm TOA},a}\times 2N_f$ matrix $\bm{F}_a$, constructed from $N_f$ sine–cosine pairs evaluated at each observation time (see, e.g., Ref.~\cite{Taylor:2021yjx} for an explicit expression). Since we are primarily interested in long-timescale correlations, the expansion is truncated after $N_f$ frequencies. In this work, we use $N_f = 30$ for modeling intrinsic red noise, and $N_f = 14$ for GWB and stochastic ULDM signals.

The first term in the parenthesis of Eq.~\eqref{eq:timing_2p} gives the timing residual power spectral density (PSD) induced by intrinsic pulsar noise, which we parametrize as
\begin{equation}
    \varphi_{a,k} = \frac{A_a^2}{12\pi^2}\frac{1}{T_{\rm obs}}\left(\frac{f_k}{{\rm yr}^{-1}}\right)^{\gamma_a}{\rm yr}^3\,,
\end{equation}
where $A_a$ and $\gamma_a$ are pulsar-dependent parameters.
The second term in the parenthesis of Eq.~\eqref{eq:timing_2p} gives the GWB contribution to the timing residual. The overlap reduction function for GWB is given by the well-known Hellings-Downs function~\cite{Hellings:1983fr}:
\begin{equation}
    \Gamma_{ab}^{\scriptscriptstyle\rm GW} = \frac{1}{2} \delta_{ab} + \frac{1}{2} - \frac{1}{4} x_{ab} + \frac{3}{2} x_{ab} \ln x_{ab}, 
\end{equation}
where $x_{ab} = ( 1- \nhat_a \cdot \nhat_b) /2$. For the GWB-induced PSD we use a similar power-law parametrization to the one used for intrinsic pulsar noise:
\begin{equation}
    \phi_k^{\scriptscriptstyle\rm GW}=\frac{A_{\scriptscriptstyle\rm GW}^2}{12\pi^2}\frac{1}{T_{\rm obs}}\left(\frac{f_k}{{\rm yr}^{-1}}\right)^{\gamma_{\scriptscriptstyle\rm GW}}{\rm yr}^3\,,
\end{equation}
the only difference is that in this case the amplitude, $A{\scriptscriptstyle\rm GW}$, and slope, $\gamma_{\scriptscriptstyle\rm GW}$, are common for all the pulsars in the array.
Finally, the last term in the parenthesis of Eq.~\eqref{eq:timing_2p} describes the ULDM contribution to the timing residual and is related to the timing residual power spectrum by 
\begin{equation}\label{eq:dm_timing}
    \Gamma_{ab}^{\scriptscriptstyle\rm DM}\phi_{k}^{\scriptscriptstyle\rm DM}=\Gamma_{ab}^{\scriptscriptstyle\rm DM}S^{\scriptscriptstyle\rm DM}(f_k) \Delta f\,,
\end{equation}
where $\Delta f=1/T_{\rm obs}$.

In theory, both the fast and slow modes of the ULDM signal could be modeled as additional red noise contributions. However, as we can see from \cref{eq:clock_fast}, the spectrum of the fast modes has support only over a narrow frequency range, $\Delta f \sim \mphi \sigma^2$, much smaller than the size of the frequency bin used in our analysis, i.e. $1/T_{\rm obs}$. Therefore, it is more convenient to treat the fast modes as a monochromatic, deterministic signal of the form:
\begin{equation}\label{coherent_dm_dt}
    \delta t^{\scriptscriptstyle\rm DM}_a(t) =  A_E \sin(2 m_\phi t + \gamma_E) + A_{P,a} \sin(2 m_\phi t + \gamma_{P,a}),     
\end{equation}
where we have split the signal into two terms: an ``Earth term" with amplitude $A_E$ and phase $\gamma_E$, and a ``pulsar term" with amplitude $A_{P,a}$ and phase $\gamma_{P,a}$. 
As we discussed in the previous section, the fast-modes are dominated by the clock and the pulsar spin signals. Specifically, the amplitudes of the Earth and pulsar terms are given by 
\begin{equation}\label{eq:amps}
    A_E =\frac{g_{\scriptscriptstyle TT}\bar\rho}{4\Mpl^2 m_\phi^3}\hat\phi_E, \,\,\,\,\,\,\, A_{P,a} = \frac{g_{I}\bar\rho}{4\Mpl^2 m_\phi^3}\hat\phi_{P,a}\,,
\end{equation}
where $g_{\scriptscriptstyle TT}$ and $g_I$ have been defined in Eq.~\eqref{eq:g_tt_main} and Eq.~\eqref{eq:g_I_main}, respectively. 
In the two equations above, $\hat\phi$ are exponentially distributed random variables with a unit scale parameter which represent fluctuations of the ULDM around its mean value (see App.~\ref{sec:gaussianity_appx} for detailed discussion). When the separation between Earth and the pulsar is smaller or comparable to the ULDM coherence length, these random variables are correlated. While, in principle, these correlations could be accurately taken into account by PTA analyses~\cite{Luu:2023rgg, Boddy:2025oxn, Li:2025xlr}, previous studies (see, for example, Refs.~\cite{NANOGrav:2023hvm, Kaplan:2022lmz}) only considered two limiting cases. (1) The so-called ``correlated" regime, in which the Earth and all the pulsars are assumed to be within an ULDM coherence patch. In this regime, we can assume that $\hat\phi_E=\hat\phi_{P,a}=\hat\phi$. (2) The ``uncorrelated" regime, in which we assume that the Earth and each of the pulsars in the array reside in a different ULDM coherence patch. In this case, we assume that $\hat\phi_E$ and each of the $\hat\phi_{P,a}$ are independent random variables. 
Due to matter effects (see next section for a detailed discussion), the pulsar signal will be relevant only in the very low mass window ($m_\phi\lesssim10^{-23}\,{\rm eV}$), where the ULDM correlation length is larger than the typical inter-pulsar distance. Therefore, in this work, we will only consider the correlated regime for deterministic ULDM signals. The phases of the Earth and pulsar terms can be written as:
\begin{equation}
    \gamma_E=\alpha_E\,,\qquad \gamma_{P,a} = 2m_\phi|\bm{x}_E-\bm{x}_{P,a}|+\alpha_{P,a}\,, 
\end{equation}
where $\alpha$ is the contribution to the phase from the underlying ULDM field configuration, while the additional term in the pulsar phase is due to the light-travel time between the pulsar and the Earth. Therefore, even in the ``correlated regime", where $\alpha_E=\alpha_{P,a}$, the phase of the Earth and pulsars signals cannot be taken to be equal because $m_\phi|\bm{x}_E-\bm{x}_{P,a}|\gg 1$ for any ULDM mass of interest. 

Given the above parametrizations for the signals contributing to the timing residuals, we can use the two-step marginalization of the timing and noise parameters described in Ref.~\cite{NANOGrav:2023icp} to obtain the likelihood function,
\begin{equation}\label{eq:pta_likelihood}
    p(\vec{\delta t}|\vec{\eta})=\frac{\exp\left(-\frac{1}{2}(\vec{\delta t}-\vec{\delta t}^{\scriptscriptstyle\rm DM})^T\vec{K}^{-1}(\vec{\delta t}-\vec{\delta t}^{\scriptscriptstyle\rm DM})\right)}{\sqrt{{\rm det}(2\pi\vec{K})}},
\end{equation}
where $\vec{\eta}$ contains all the model parameters including the red noise and ULDM parameters, and $\vec{K}=\vec{D}+\vec{F}\vec{\Phi}\vec{F}^T$.
Here, $\vec{D}=\vec{N}+\vec{MEM}^T$, where $\vec{M}$ is a $N_{\rm TOA}\times m$ matrix, which contains the partial derivatives of the TOAs with respect to the timing-model parameters (evaluated at the best-fit values), and $\vec{E}=\langle\vec{\epsilon\epsilon}^T\rangle$ is set to be a diagonal matrix of very large values ($10^{40}$), which effectively means that we assume flat priors for the parameters in $\vec{\epsilon}$. In the second term of $\bm{K}$ we have also introduced
\begin{equation}
    \bm{\Phi}_{ab}\equiv \delta_{ab}\bm{\varphi}_a+\Gamma_{ab}^{\scriptscriptstyle\rm GW}\bm{\phi}^{\scriptscriptstyle\rm GW}+\Gamma_{ab}^{\scriptscriptstyle\rm DM}\bm{\phi}^{\scriptscriptstyle\rm DM}\,.
\end{equation}

\subsection{Bayesian Analysis}
We implement the search for the ULDM signals discussed in the previous sections by using the \texttt{PTArcade} package~\cite{Mitridate:2023oar}. This package allows us to easily implement the PTA likelihood defined in Eq.~\eqref{eq:pta_likelihood} into \texttt{enterprise}~\cite{ENTERPRISE}, and perform Markov Chain Monte Carlo (MCMC) runs to estimate the posterior distributions of the timing model parameters. The list of timing model parameters considered in this work is summarized, together with their priors, in Table~\ref{tab:priors_stoch} and Table~\ref{tab:priors_det}.\footnote{\texttt{PTArcade} model files for the ULDM signals discussed in this work can be found in \href{https://zenodo.org/records/17341715}{this Zenodo repository.}} Following the standard practice, in our ULDM searches we keep white-noise parameters fixed. In the analysis of the NANOGrav 12.5-year data set, we fix these parameters to their maximum likelihood value extracted from single-pulsar analyses; while for the analysis of the mock 30-year data set, we fix the white-noise parameters to their injected values.

In the absence of an ULDM signal, we can then marginalize over timing model parameters except for the DM mass and the relevant DM coupling to matter. The constraints on the relevant DM coupling in a given mass-bin are then set the $95^{\rm th}$ percentile of the posterior distribution in that bin. 

\subsection{Data Sets}\label{subsec:mock}
In this work, we use two sets of PTA data: the NANOGrav 12.5-year data set, which we use to test and calibrate our search on real data set; and a mock PTA data set, which we use to derive projections for the sensitivity of future PTAs to the signals considered in this study. Details about the NANOGrav 12.5-year data set can be found in Ref.~\cite{NANOGrav:2020gpb}.

To generate the mock data, we use \texttt{libstempo}~\cite{2020ascl.soft02017V}, a Python wrapper for \texttt{TEMPO2}~\cite{Edwards:2006zg, Hobbs:2006cd}. We start from a core catalog comprising 67 pulsars from the NANOGrav 15-year data set~\cite{NANOGrav:2023hde}, for which we assume a 15-year observational baseline. This catalog is then expanded by adding 33 pulsars every five years, until a total observing time of 30 years is reached.

The final data set contains 166 pulsars randomly placed in the sky. Of these, 67 pulsars have a 30-year observational baseline, while the remaining 99 are divided into three groups with baselines of 15, 10, and 5 years, respectively. For all pulsars, we assume a nominal observing cadence of three weeks, with small random fluctuations applied. Each TOA is assigned an uncertainty, $\sigma_{a,i}$, drawn from a normal distribution with mean and standard deviation specified in Table~\ref{tab:noise_dist}. The synthetic TOAs are injected with both white and red noise, as well as a GWB signal. These contributions are modeled using the statistical framework described in Sec.~\ref{subsec:pta_likelihood}. Noise parameters for each pulsar are randomly sampled from the distributions in Table~\ref{tab:noise_dist}, while the GWB parameters are fixed to $\log_{10}A_{\rm GWB} = -14.2$ and $\gamma_{\rm GWB} = 3.2$, close to the values observed in the most recent NANOGrav data set~\cite{NANOGrav:2023gor}. These parameters were chosen to reflect the noise properties of current PTAs. Although the ULDM correlation function partially breaks the degeneracy with intrinsic pulsar red noise and the GWB, increasing either the intrinsic red-noise level or the GWB amplitude will reduce sensitivity to the ULDM singal, and vice versa.

\begin{table}[t]
\bgroup
\renewcommand{\arraystretch}{1.5}
\setlength\tabcolsep{3.pt}
{\footnotesize
	\begin{tabular}{ccc}
		\toprule
		\textbf{Parameter}	                       &   \textbf{Description}                 &   \textbf{Distribution}   \\ \hline
		$\sigma_{a,i}$                             &   TOA uncertainty                      &   $\mathcal{N}(400\,{\rm ns} , 200\,{\rm ns})$    \\ \midrule
		$\mathcal{F}_{a}$                          &   EFAC                                 &   $\mathcal{N}(1, 0.05)$    \\   
		$\log_{10}\mathcal{Q}_{a}$                 &   EQUAD                                &   $\mathcal{N}(-8.5, 1)$    \\ \midrule
		$\log_{10} A_a$                            &   intrinsic red-noise amplitude        &   $\mathcal{N}(-16, 1)$    \\
		$\gamma_a$                                 &   intrinsic red-noise spectral index   &   $\mathcal{U}(1, 5)$        \\ 
        \bottomrule
	\end{tabular}
}
\egroup
\caption{Probability distributions used to generate values of noise parameters. In the last column, $\mathcal{N}(\mu,\,\sigma)$ indicates a normal distribution with mean $\mu$ and standard deviation $\sigma$.\label{tab:noise_dist}}
\end{table}

\section{Results}\label{sec:results}

\subsection{Dilaton-like Scalar}
\begin{figure*}[t]
\centering
\includegraphics[width=\textwidth]{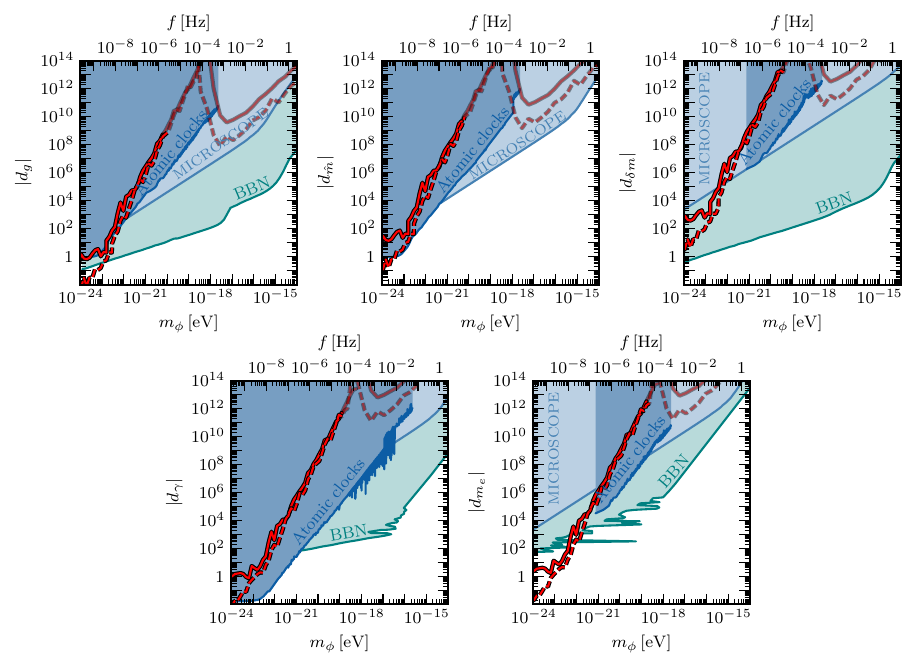}
\caption{Projected sensitivities to the dilaton couplings $d_i$ from NANOGrav 12.5-year (solid) and simulated data set of total 30-year observation baseline (dashed). Fully opaque lines denote unscreened sensitivities, whereas semi-transparent red lines indicate sensitivities screened by matter effect. The sensitivities at masses around $10^{-17}\,{\rm eV}$ arise from the stochastic fluctuations at $\omega \lesssim m_\phi \sigma^2$, while the constraints at lower masses arise from coherent fluctuations at $\omega = 2m_\phi$. Other constraints --- from MICROSCOPE equivalence principle tests~\cite{Hees:2018fpg,MICROSCOPE:2019jix,MICROSCOPE:2022doy}, BBN~\cite{Stadnik:2015kia,Sibiryakov:2020eir,Bouley:2022eer}, and atomic clocks~\cite{Hees:2016gop, Kennedy:2020bac, BACON:2020ubh, Oswald:2021vtc, Sherrill:2023zah, Filzinger:2023zrs} --- are shown.}
\label{fig:main}
\end{figure*}

We present the main results of our analysis in Fig.~\ref{fig:main}. Each panel of this figure shows the limits on the dilaton couplings $d_i$ from the analyses of the NANOGrav 12.5-year data set (red solid)~\cite{NANOGrav:2020gpb} and the simulated data set with the total 30-year time baseline (red dashed)~\cite{hyungjin_2024_10534322}. We analyze both coherent and stochastic signals. The coherent signal search covers the lower mass range, $10^{-24}\eV\textrm{--}10^{-22}\eV$, while the stochastic signal search probes a mass range roughly six orders of magnitude higher than that of the coherent search, i.e. $10^{-18} \eV\textrm{--}10^{-16}\eV$. Other limits, such as those from tests of the equivalence principle~\cite{Hees:2018fpg, MICROSCOPE:2019jix, MICROSCOPE:2022doy}, big bang nucleosynthesis (BBN)~\cite{Stadnik:2015kia,Sibiryakov:2020eir,Bouley:2022eer}, and atomic clocks~\cite{VanTilburg:2015oza, Hees:2016gop, Kennedy:2020bac, BACON:2020ubh, Oswald:2021vtc, Sherrill:2023zah, Filzinger:2023zrs}, are overlaid. All results shown in the figure are obtained by assuming that only one dilaton coupling is nonzero and the other is zero, and ignoring the gravitational contribution to the timing residuals produced by ULDM oscillations. Note that the shown limit from the MICROSCOPE equivalence principle test assumes the positive sign of the dilaton coupling such that the interaction between ULDM and the Earth is repulsive; for the negative sign where the interaction becomes attractive, the resulting limit will exhibit narrow resonance features~\cite{Gue:2025nxq, Delaunay:2025pho} due to the matter effect, which will be discussed shortly. In addition, a static scalar field profile could be sourced in the case of attractive interaction, and in such cases, stellar objects such as white dwarfs could be used to place limits on the existence of quadratically ultralight scalar fields~\cite{Bartnick:2025lbg}. 

The upper limits on the dilaton couplings shown in Fig.~\ref{fig:main} should be taken with some care. Quadratic couplings can significantly change the mass of dark matter within the Sun, Earth, and pulsars, which in turn modifies the dark matter field profile around these objects~\cite{Hees:2018fpg, Banerjee:2022sqg, Banerjee:2025dlo, delCastillo:2025rbr, Gan:2025nlu}. For sufficiently large dilaton couplings, this matter effect tends to suppress the field value within these objects, thereby suppressing dark-matter-induced signals. In the region of the ULDM parameter space where this happens, we have decreased the opacity of our constraint line. For the coherent signal, a large portion of our constraints is not affected by this matter effect; however, the entirety of the stochastic signal is impacted by this effect. We will discuss matter effects in more detail later in this section and in App.~\ref{sec:matter_effect_appx}.

\subsection{Light QCD Axion}
QCD axion models are one example of theories that predicts quadratic couplings between ultralight scalars and the Standard Model fields. By a QCD axion, we refer to a pseudo Nambu-Goldstone boson that couples to the gluon field strength tensor, $G_{\mu\nu}^a$, as
\begin{equation}
{\cal L} = \frac{g^2_3}{32\pi^2} \frac{\phi}{\fphi} G_{\mu\nu}^a \widetilde G^{a\mu\nu}\,,
\end{equation}
where $\widetilde G_{\mu\nu}^a$ is the dual of $G_{\mu\nu}^a$, and $\fphi$ is the axion decay constant. Below the QCD scale, the strong sector confines, explicitly breaking the underlying axion shift symmetry. Consequently, the axion develops quadratic couplings to mesons and hadrons~\cite{Kim:2022ype}, as well as to the photon and the electron fields via loop corrections~\cite{Beadle:2023flm, Kim:2023pvt,Gan:2025nlu,Bai:2025yxm}.

We therefore expect QCD axion dark matter to give rise to PTA signals similar to the ones discussed in previous sections. Specifically, it will lead to both coherent and stochastic signals that mainly stem from its coupling to the nucleon mass:
\begin{equation}\label{axion_quad}
    {\cal L} = 
    - C_N \frac{\phi^2}{2 \fphi^2} m_N \bar N N , 
\end{equation}
where $N = (p,n)$ denotes the nucleon field, and $C_N \sim - 10^{-2}$~\cite{Ubaldi:2008nf,Kim:2022ype}. For any object composed mainly of nucleons, the axion coupling strength for the Doppler is given by
\begin{equation}
    g_N = C_N \frac{\Mpl^2}{\fphi^2}\,. 
\end{equation}
At the same time, axions will also lead to the clock signals in the timing observations. Assuming that the frequency of Terrestrial Time depends mainly on the Cs frequency standard, we find that the coupling strength for the clock signal is~\cite{Kim:2022ype}
\begin{equation}
    g_{\scriptscriptstyle\rm TT} \simeq -0.01 \frac{\Mpl^2}{\fphi^2}\,,
\end{equation}
where we have only considered the effects due to the variation of the $g$-factor and the proton mass, and ignored those from the variation of the fine structure constant and the electron mass as they are subleading. 

\begin{figure}[t]
\centering
\includegraphics[width=\linewidth]{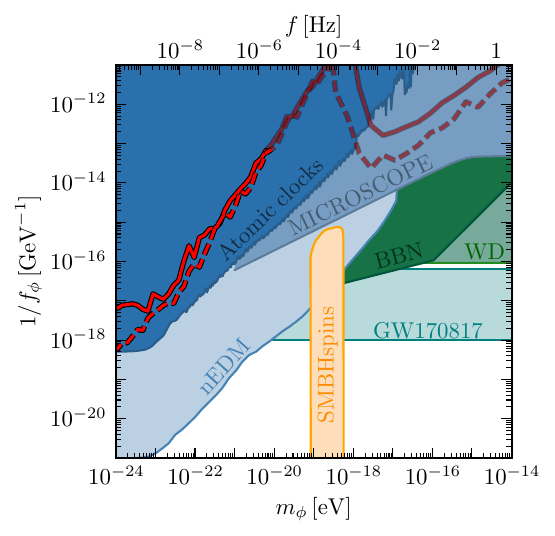}
\caption{Projected sensitivities to the light QCD axion from the NANOGrav 12.5-year (solid) and simulated data set with a total 30-year observation baseline (dashed). Fully opaque lines denote unscreened sensitivities, whereas semi-transparent red lines indicate screened sensitivities. The limits at low mass arise from searches for coherent fluctuations at $\omega = 2m_\phi$, while the sensitivities at high mass arise from searches for stochastic fluctuations at $\omega \lesssim m_\phi \sigma^2$. Other limits, e.g. those from MICROSCOPE~(App.~\ref{subsec:MICROSCOPE_Axion} and Ref.~\cite{Gue:2025nxq}), atomic clock comparison tests~\cite{Hees:2016gop, Kennedy:2020bac, BACON:2020ubh, Oswald:2021vtc, Sherrill:2023zah, Filzinger:2023zrs}, oscillating nEDM~\cite{Abel:2017rtm}, SMBH spin-down~\cite{Arvanitaki:2014wva, Baryakhtar:2020gao, Hoof:2024quk}, GW170817~\cite{Zhang:2021mks}, BBN~\cite{Blum:2014vsa}, and white dwarf mass-radius relation~\cite{Balkin:2022qer}, are also shown.}
\label{fig:main_axion}
\end{figure}

From the null observation of the ULDM-induced coherent and stochastic signals in NANOGrav 12.5-year and simulated timing data sets, we are able to place constraints and projections in the plane of axion mass and its decay constant, shown in Fig.~\ref{fig:main_axion}. In this figure, we consider the axion mass as if it were a free parameter. While this assumption is not justified for minimal QCD axion models, where the axion mass is fixed as $m_\phi^2 \fphi^2 \simeq m_\pi^2 f_\pi^2$ with $m_\pi$ and $f_\pi$ being the pion mass and decay constant, it is still possible to reach the parameter space above this line in a technically natural way at the price of further model building~\cite{Hook:2018jle,DiLuzio:2021pxd,DiLuzio:2021gos}. In Fig.~\ref{fig:main_axion}, we also overlay other constraints such as those from atomic clock comparison tests and molecular spectroscopy~\cite{VanTilburg:2015oza, Hees:2016gop, Kennedy:2020bac, BACON:2020ubh, Sherrill:2023zah, Filzinger:2023zrs}, oscillating neutron electric dipole moment~(nEDM)~\cite{Abel:2017rtm}, supermassive black hole~(SMBH) spin-down~\cite{Arvanitaki:2014wva, Baryakhtar:2020gao, Hoof:2024quk},
the GW170817 signal from neutron star mergers~\cite{Zhang:2021mks}, white dwarf mass-radius relation~\cite{Balkin:2022qer}, BBN~\cite{Blum:2014vsa}, and MICROSCOPE equivalence principle tests~(see App.~\ref{subsec:MICROSCOPE_Axion} and Ref.~\cite{Gue:2025nxq} for a detailed discussion).

As in the dilaton case, matter effects can suppress axion-induced signals. In Fig.~\ref{fig:main_axion}, we have decreased the opacity of the constraints lines in the region where matter effects suppress the PTA signal.

\subsection{Matter Effect}\label{sec:matter_effect_main}

Quadratic interactions affect the evolution of the dark matter field near astrophysical objects. As a consequence, the dark matter field profile is distorted from a simple plane wave, and this distortion changes the expected ULDM-induced timing signals, computed in the earlier sections with the assumption that the dark matter field consists of a linear combination of plane waves. This phenomenon is referred to as the matter effect. Such effects have previously been investigated in Refs.~\cite{Day:2023mkb,Banerjee:2025dlo,delCastillo:2025rbr,Gan:2025nlu,Burrage:2025grx}. See Ref.~\cite{Gan:2025nlu} for a detailed classification and discussions covering both the low- and high-momentum regimes. Building upon the formalism developed in these studies, we investigate how the matter effect modifies the PTA measurements. 

The matter effect arises from the difference between the in-medium and vacuum masses of dark matter. As an illustration, consider a quadratic coupling
\begin{equation}\label{eq:L_OSM_main}
    {\cal L} = - \frac{\phi^2}{2\Mpl^2} \, \bm{d} \cdot \vecosm,
\end{equation}
where $\vecosm$ represents a set of Standard Model operators, such as $m_N \bar{N} N$ or those appearing in the Lagrangian of Eq.~(\ref{eq:lagrangian}). In a finite-density environment, $\vecosm$ can be replaced by its ensemble average, $\vecosm \rightarrow \langle \vecosm \rangle = \bm{Q} \, \rho$, where $\bm{Q}$ and $\rho$ are the charge and the density of the dense object under consideration. In this case, the dark matter mass acquires an additional contribution:
\begin{equation}\label{eq:msq_tot_eq_min}
    m_\tot^2({\boldsymbol x}) = m_\phi^2 + \Delta m^2({\boldsymbol x})\,
\end{equation}
with
\begin{equation}\label{eq:Del_mMSq_main}
    \Delta m^2({\boldsymbol x}) = \frac{g \, \rho({\boldsymbol x})}{\Mpl^2}
\end{equation}
denoting the difference between the squared masses in medium and in vacuum. 
Here, $g = \bm{d} \cdot \bm{Q}$ denotes the effective coupling between the scalar and the ordinary matter, and $m_\tot^2({\boldsymbol x})$ is the total in-medium mass squared.  This medium-induced mass correction leads to a non-trivial evolution of the ULDM field in proximity of massive objects.

A straightforward way to study the field evolution is to solve the Klein–Gordon equation with a spatially dependent mass. For ULDM, it is convenient to work in the non-relativistic limit. In this limit, a mode with incident momentum ${\boldsymbol k}$ can be decomposed as
\begin{equation}
    \phi_{\boldsymbol k}(t,\boldsymbol x) = \frac{1}{\sqrt{2m_\phi V}} a_{\boldsymbol k} \psi_{\boldsymbol k}(\boldsymbol x) e^{- i \omega_{\boldsymbol k} t} 
    + {\rm h.c.},
\end{equation}
with $\omega_{\boldsymbol k}^2 = |{\boldsymbol k}|^2 + \mphi^2$ and the fiducial normalization volume $V$. The Klein-Gordon equation reduces to the time-independent Schr{\"o}dinger equation:
\begin{equation}\label{eq:schrodinger_matter}
    E_{\boldsymbol k} \psi_{\boldsymbol k}(\boldsymbol x)
    = 
    \left[
    - \frac{\vecnabla^2}{2m_\phi}
    + V_\text{eff}({\boldsymbol x})
        \right] \psi_{\boldsymbol k}(\boldsymbol x).
\end{equation}
Here, $E_{\boldsymbol k} = |{\boldsymbol k}|^2/2m_\phi$ denotes the kinetic energy, and 
\begin{equation}
    V_\text{eff}({\boldsymbol x}) = \frac{\Delta m^2({\boldsymbol x})}{2m_\phi}
\end{equation}
is the effective potential experienced by the ULDM field when propagating through a massive object. In the limit $\Delta m \, R \to 0$, where $R$ is the radius of the astrophysical object under consideration, the effective potential contribution vanishes and the initial plane wave remains unchanged when propagating through the massive object; this corresponds to the regime in which the ULDM-induced PTA signals are not impacted by matter effects. By contrast, a nonzero mass difference induces scattering of the wave around the object. In particular, in the limit $\Delta m \, R \to \infty$, the target behaves as a hard sphere, forcing the field outward. In this regime, the ULDM-induced PTA signals are strongly suppressed.

Between these two regimes, matter effects can be included into the ULDM-induced PTA signals by accounting for the field distortion around relevant astrophysical objects. As shown in App.~\ref{sec:matter_effect_appx}, the timing residuals can be expressed as
\begin{equation}\label{matter_timing_signal}
    [\delta t_a(t)]_{i} =  {\cal A}_{i} (y)\times [ \delta t_a(t) ]_{i, 0},    
\end{equation}
where the dimensionless parameter
\begin{equation}\label{eq:y_parameter}
    y = \Delta m R    
\end{equation}
characterizes the strength of the quadratic coupling. Here, $[\delta t_a(t)]_{i,0}$ denotes the timing residual in the absence of matter effects, $\mathcal{A}_i(y)$ is a form factor that encodes the impact of matter effects, and the index $i=\dop,\,\clk,\,\psr$ labels the Doppler, clock, and pulsar spin signals, respectively. Throughout this work, we take the low-momentum limit $kR \ll 1$, which is valid across all mass ranges considered. In this limit, the corresponding form factors ${\cal A}_i(y)$ are given by
\begin{align}
\label{eq:form_factor_dop}
    {\cal A}_\dop(y) &\simeq  3 \left(\frac{y - \tanh y}{y^3}\right), \\
    \label{eq:form_factor_clk}
    {\cal A}_\clk(y) &\simeq \frac{\tanh^2 y}{y^2},\\
    \label{eq:form_factor_psr}
    {\cal A}_\psr(y) &\simeq \frac{3}{2} \left( \frac{\tanh^2 y}{y^2} - \frac{y-\tanh y}{y^3} \right).
\end{align}
Their behavior is illustrated in Fig.~\ref{fig:form_factors} in Appendix. Note that all form factors converge to unity in the regime $y \lesssim 1$. Each form factor accounts for the following matter effect: the Doppler form factor accounts for the modified bulk average of the field gradient, the clock form factor accounts for the modified field value at the surface of Earth, and the pulsar-spin form factor accounts for the modified bulk average of the pulsar radius and mass. Detailed derivations are provided in App.~\ref{sec:matter_effect_appx}.\footnote{In the above discussion, we have assumed that the potential is repulsive, $V_{\rm eff} = \Delta m^2 / 2m_\phi > 0$. The same discussion applies to the case of an attractive potential $V_{\rm eff} = \Delta m^2/2m_\phi < 0$. In this case, the form factors, which can be obtained from Eqs.~\eqref{eq:form_factor_dop}--\eqref{eq:form_factor_psr} by analytic continuation $y \rightarrow -iy$, exhibit narrow resonance features. For a large $y$, all form factors scale as $1/y^2$, leading to a suppression of the PTA signals.}

The form factors in Eqs.~(\ref{eq:form_factor_dop})$\,\textrm{--}\,$(\ref{eq:form_factor_psr}) indicate that the matter effect becomes important when
$$
|y| \gtrsim 1 \quad \Longleftrightarrow \quad |g| \gtrsim \frac{\Mpl^2}{\rho R^2}.
$$ 
In this limit, the Doppler and clock form factors scale as ${\cal A}(y) \propto 1 /y^2 \propto 1 / g$. The pulsar spin form factor scales as ${\cal A}(y) \propto 1 /y^3 \propto 1 / g^{3/2}$. Since the ULDM signal without matter effect scales as $[\delta t_a(t)]_0 \propto g$, we see that the Doppler and clock signals saturate to a constant value in this limit, independent of the coupling strength; meanwhile, the pulsar-spin signal becomes increasingly suppressed as the coupling grows stronger. Therefore, ULDM signals are {\it screened} once the coupling exceeds the critical strength, which can be expressed as follows:
\begin{align}\label{eq:matter_line_main}
    & \text{Doppler}: \quad & & |g_\odot|
    \gtrsim 4\times 10^5, \\
    & \text{Clock}: \quad & & |g_\oplus|
    \gtrsim  10^9, \\
    & \text{Pulsar Spin}: \quad & & |g_{\rm psr}| 
    \gtrsim 5 . 
\end{align}
In other words, physically meaningful constraints can only be achieved when the sensitivities from pulsar timing exceed the above critical thresholds. To show physically meaningful PTA limits on dilatons and axions, we plot a portion of the sensitivity curves in Fig.~\ref{fig:main} and Fig.~\ref{fig:main_axion} as semi-transparent red lines when they do not exceed the above threshold values. In the search for coherent signals, the clock signal is mostly unaffected, while the pulsar-spin signal is screened by the matter effect except in a very low-mass range $m_\phi \lesssim 10^{-23}\eV$. In the search for stochastic signals, the Doppler signal is screened across the entire mass range.

\subsection{Gaussianity}\label{subsec:gaussianity}
The PTA likelihood \eqref{eq:pta_likelihood} is derived by assuming that all time-correlated signals in the timing residuals are Gaussian processes. For the case of the ULDM signals considered in this work, it is not immediately clear if this assumption is justified. Indeed, since ULDM signals are proportional to the square of the underlying Gaussian field, i.e. $(\delta t)_{\rm DM}\propto\phi^2$, they are generally expected to be non-Gaussian. In this section we will argue that despite the quadratic nature of the signal, the Gaussian assumption is still justified.

To illustrate this point, we start by expanding the timing residual induced by the ULDM signals as 
\begin{equation}
    [\delta t(t)]^{\rm DM}  = \sum_{k=1}^{N_f}
    \Big[
    s_k^{\rm DM} \sin(2\pi f_k t) + c_k^{\rm DM} \cos(2\pi f_kt)
    \Big],
\end{equation}
where $f_k = k /T$ and $T$ is the total time baseline. Each of the expansion coefficients, $(s_1, \, c_1, \, \cdots, \, s_{N_f}, \, c_{N_f})^{\rm DM}$, is a random variable, whose probability distribution is specified by the underlying ultralight dark matter distribution. The degree of non-Gaussianity may be estimated by comparing the $2$- and $4$-point functions of these coefficients (note that the odd-point function vanishes as we assume an isotropic velocity distribution). In particular, we can compute their Kurtosis as
\begin{equation}
    {\rm Kurtosis}(c_k)
    = \frac{\langle c_k^4 \rangle}{\langle c_k^2 \rangle^2}
    \simeq
    3\left[ 1 + \frac{9}{64} \frac{2 \pi \Delta f}{m_\phi\sigma^2} \right]
\end{equation}
where $\Delta f = 1 / T$ and we drop the superscript DM for brevity. The same result holds for the Fourier sine coefficients, $s_k$. See App.~\ref{sec:gaussianity_appx} for details. If $c_k$ were Gaussian, their Kurtosis should be $3$. For the stochastic ULDM signal, the deviation from the Gaussian expectation scales as $\Delta f/ m_\phi\sigma^2$, which is always small in the regime of validity of the ULDM stochastic analysis. Therefore, we conclude that the stochastic ULDM signal can be safely modeled as a Gaussian process.

\subsection{Time and Frequency Standard}
Pulsar timing array collaborations record and analyze timing residuals using the Terrestrial Time~\cite{NANOGrav:2023hde, Zic:2023gta, EPTA:2023sfo}. In this work, we assume that the ULDM-induced shift in the frequency of Terrestrial Time is identical to that of Cs frequency standard, i.e. $\delta f_{\scriptscriptstyle\rm TT}/ f_{\scriptscriptstyle\rm TT} = \delta f_{\rm Cs} / f_{\rm Cs}$. To investigate how Terrestrial Time responds to the underlying dark matter fluctuations, one must understand how it is constructed in the first place. 

Terrestrial Time is constructed on the basis of international atomic time (TAI). To construct TAI, the Bureau International des Poids et Mesures (BIPM) collects clock comparison data of $\gtrsim 400$ atomic clocks located in laboratories spread throughout the world. Currently, one half of the clock ensemble consists of commercial cesium clocks and the other half of hydrogen masers. BIPM then computes a free atomic scale (EAL) from a weighted average of individual clock readings. The weights are given such that the resulting EAL achieves the maximum stability. Currently, around $90\%$ weights are assigned to hydrogen masers, and hence the frequency of EAL is expected to be sensitive to fluctuations in hydrogen maser frequency. See Ref.~\cite{2019Metro..56d2001P} for a review.

The frequency of EAL is then compared with that of primary frequency standards. These primary standards are cesium fountain clocks. From the measurement of frequency comparison between EAL and primary standards, one estimates the frequency offset between EAL and a collection of primary frequency standards~\cite{1977Metro..13...87A}. With this estimated offset, the frequency of EAL is steered to conform to the definition of SI second, which is based on the ground state hyperfine transition in cesium atom. The resulting time-scale unit is TAI. Terrestrial Time is obtained by adding $32.183\,{\rm s}$ to TAI. We therefore expect that the frequency of Terrestrial Time follows mostly that of cesium atom as a result of frequency steering.

Since the frequency steering is applied on a monthly basis, the terrestrial time may respond differently to the coherent ULDM signal when its mass is larger than $1/{\rm month}$. In this case, the coherent signal oscillates rapidly during the period over which the steering correction is computed, and as a result, ULDM signal averages out in the correction. The resulting frequency standard will respond to dark matter as if steering correction had not been applied, suggesting $\delta f_{\rm TT} /f_{\rm TT} = \delta f_{\rm EAL} / f_{\rm EAL}$ for $m_\phi > 1/ {\rm month}$. Since the EAL is constructed from the ensemble of atomic clocks, and its composition varies over time, the response or sensitivity coefficient varies over time as well. For the current analysis, we still choose the sensitivity coefficient of cesium atom for simplicity.

\section{Conclusions}\label{sec:conclusions}
We have examined the sensitivities of current and future pulsar timing arrays to signals of ULDM candidates that quadratically couple to the Standard Model. We show that, due to the quadratic nature of interaction, two types of signals appear simultaneously in PTA data.
The first, which we call coherent signal, is characterized by an angular frequency $\omega = 2m_\phi$. The second, which we call stochastic signal, is relevant for angular frequencies $\omega \lesssim m_\phi \sigma^2$. The distinctive frequency components of these signals allow us to probe a widely separated mass range, spanning from the fuzzy dark matter regime, $m_\phi = 10^{-24}\eV \, \textrm{--} \, 10^{-22}\eV$, to a heavier mass range, $m_\phi = 10^{-18}\eV  \, \textrm{--} \, 10^{-16}\eV$. We show how current PTA data sets are already placing significant constraints in the low mass range, while for the higher mass range, due to the matter effect, they do not yet provide physically meaningful constraints.

We apply the same analysis for the light QCD axion model. Due to its coupling to gluon and an explicit shift symmetry breaking from the strong dynamics, axion generally develops quadratic couplings to the nuclear sector at low-energy scales. We project the PTA sensitivities in the axion mass-decay constant plane. The same conclusion holds in this case; for the stochastic signal, due to matter effects, current PTAs do not place physically meaningful constraints, while for the coherent signal, the limits from PTA analyses compete with existing terrestrial bounds such as the ones from atomic clock comparison tests. 

\bigskip 

\acknowledgments
We thank Alessandro Lenoci for providing us relevant data for the limits presented in a light QCD axion plot. We also thank Abhishek Banerjee, Yifan Chen, Konstantin Springmann, and Bingrong Yu for reading the draft of this manuscript and providing us useful comments and suggestions. This work was supported by the Deutsche Forschungsgemeinschaft under Germany's Excellence Strategy - EXC 2121 Quantum Universe - 390833306.
\bigskip

\appendix

\section{Gaussianity}\label{sec:gaussianity_appx}

In this appendix, we carefully examine statistical properties of the coherent and stochastic dark matter signals. We focus on clock and Doppler signals, each of which is most relevant for the coherent and stochastic signal search, respectively. The response of the pulsar spin signal with respect to ULDM is identical to that of clock signal, and hence, the discussion can be straightforwardly generalized.

Before proceeding to the analysis, we review the properties of the ULDM Gaussian random field. We expand this field in the non-relativistic regime as
\begin{align}
\phi(t, \boldsymbol x)
= \sum_{\boldsymbol k}
\frac{1}{\sqrt{2 \mphi V}}
\Big[ 
a_{\boldsymbol k} e^{- i k \cdot x}
+ a_{\boldsymbol k}^* e^{i k \cdot x}
\Big]\,,
\label{field_expansion_free_space}
\end{align}
where $V$ is the normalization volume in which the field is expanded. We treat $a_{\boldsymbol k}$ and $a^*_{\boldsymbol k}$ as classical random variables that can be expanded as
\begin{align}
a_{\boldsymbol k}
= r_{\boldsymbol k} e^{i \theta_{\boldsymbol k}},
\end{align}
where the amplitude, $r_{\bm{k}}$, and phase, $\theta_{\bm k}$, follow the Rayleigh and uniform distributions, respectively~\cite{Derevianko:2016vpm, Foster:2017hbq, Centers:2019dyn, Kim:2021yyo, Cheong:2024ose}:
\begin{align}
\mathcal{P}(r_{\boldsymbol k}) 
&= \frac{2 r_{\boldsymbol k}}{f_{\boldsymbol k}}
\exp\left(
- \frac{r_{\boldsymbol k}^2}{f_{\boldsymbol k}^2}
\right), 
\\
\mathcal{P}(\theta_{\boldsymbol k}) 
&= 
\frac{1}{2\pi}  . 
\end{align}
The parameter $f_{\boldsymbol k}$ is understood as a momentum distribution of dark matter. In the continuum limit, it is normalized as
\begin{align}
\frac{\bar \rho}{m_\phi} = \int \frac{d^3 {\boldsymbol k}}{(2\pi)^3} f(\boldsymbol k) 
= \int d^3 {\boldsymbol v} \, f(\boldsymbol v),
\end{align}
where $\bar\rho$ is the mean dark-matter density. The above expression defines the velocity distribution $f(\boldsymbol v)$.

The random variable $a_{\boldsymbol k}$ can also be decomposed into real and imaginary parts:
\begin{equation}
a_{\boldsymbol k}= X_{\boldsymbol k} + i Y_{\boldsymbol k},
\end{equation}
In this expansion, the variables $(X_{\boldsymbol k}, Y_{\boldsymbol k})$ are independent and identically distributed random variables, following a zero-mean normal distribution, i.e.,
\begin{equation}
X_{\boldsymbol k}, \, Y_{\boldsymbol k} \sim {\cal N}(0, f_{\boldsymbol k}/2).
\end{equation}
Here, ${\cal N}(\mu, \sigma^2)$ represents the normal distribution with mean $\mu$ and variance $\sigma^2$.

\subsection{Coherent Signal}
The coherent signal produced by ULDM-induced clock shifts is given by
\begin{align*}
\delta t_a (t) 
&= g_{\scriptscriptstyle\rm TT} \int^t dt' \, \varphi^2(t') \\
&= \frac{g_{\scriptscriptstyle\rm TT}}{4 \Mpl^2 m_\phi^{2} }
   \mathcal{R}^2 \sin \!\left( 2 m_\phi t - 2 \Theta \right) \, .
\end{align*}
where used the non-relativistic approximation $\omega = m_\phi$, and implicitly defined ${\cal R}$ and $\Theta$ via 
\begin{align}
\frac{1}{\sqrt{V}} \sum_{\boldsymbol k} a_{\boldsymbol k}
= {\cal R} e^{i\Theta} . 
\label{RTheta}
\end{align}

To find how ${\cal R}$ and $\Theta$ are distributed, we decompose them into the real and imaginary part:
\begin{align}
\frac{1}{\sqrt{V}} \sum_{\boldsymbol k} a_{\boldsymbol k}
= X + i Y , 
\end{align}
where 
\begin{align}
X 
&= \frac{1}{\sqrt{V}} \sum_{\boldsymbol k} X_{\boldsymbol k},
\\
Y &= \frac{1}{\sqrt{V}} \sum_{\boldsymbol k} Y_{\boldsymbol k}
.
\end{align}
The statistical distribution of $X$ and $Y$ can be found by computing their characteristic function. The characteristic function for $X$ is
\begin{align}
\varphi_X(t) = 
\langle e^{i t X} \rangle
&= \prod_{\boldsymbol k} 
\left\langle \exp\left[ \frac{i t X_{\boldsymbol k}}{\sqrt{V}} \right] 
\right\rangle
\nonumber\\
&= 
\exp\Big[ 
-\frac{t^2}{4V}
\sum_{\boldsymbol k}
f_{\boldsymbol k}
\Big] . 
\end{align}
Here we used the fact that each $X_{\boldsymbol k}$ is independent to each other. The result is identical to that of a zero-mean normal random variable with variance,
\begin{align}
\sigma^2 
= \frac{1}{V} \sum_{\boldsymbol k} \frac{f_{\boldsymbol k}}{2}
= \int \frac{d^3 {\boldsymbol k}}{(2\pi)^3} \frac{f(\boldsymbol k)}{2}
= \frac{\bar\rho}{2m_\phi}\,,
\end{align}
from which one finds
$$ X,\,Y \sim {\cal N}(0, \bar\rho/2m_\phi).$$ 
This implies that $\cal R$ and $\Theta$ follow a Rayleigh and uniform distribution, respectively:
\begin{align}
\mathcal{P}( {\cal R} ) 
&= \frac{2 {\cal R}}{(\bar \rho/m_\phi)}
\exp\left[
- \frac{{\cal R}^2}{(\bar \rho / m_\phi)}
\right],
\label{p(R)}
\\
\mathcal{P}(\Theta)
&= \frac{1}{2\pi}. 
\label{p(Theta)}
\end{align}
Once we rescale ${\cal R} \to {\cal R} \sqrt{\bar\rho / m_\phi}$, we arrive at Eq.~\eqref{coherent_dm_dt} that we use for the coherent signal search. In this case, the square of the rescaled amplitude ${\cal R}^2$ follows the exponential distribution with the unit scale parameter.

\subsection{Stochastic Signal}
The timing residual of the ULDM-induced Doppler signals is:
\begin{align}
    & \quad \,\,\, \widetilde{\delta t}_a (f) =
    \int \frac{d^3 {\boldsymbol k}}{(2\pi)^3} 
    {\cal K}_a(f, \boldsymbol k) \widetilde{\varphi^2}(f, \boldsymbol k)
    \nonumber\\
    &= 
    \frac{1}{2\Mpl^2 m_\phi}
    \frac{1}{V}
    \sum_{1,2} a_1 a_2^*
    {\cal K}_a(f,\boldsymbol k_1 - \boldsymbol k_2)
    \delta(f - f_1 + f_2).
\end{align}
The sum runs over all possible $\boldsymbol k_1$ and $\boldsymbol k_2$ with $\boldsymbol k_1 \neq \boldsymbol k_2$. In the following, we ignore the pulsar term, i.e. ${\cal K}_a(f,\boldsymbol k) \simeq - [g_\odot / (2\pi f)^2] (i \boldsymbol k \cdot \hat{\boldsymbol n}_a)$. 

To check the Gaussianity of this random variable, we compute the $n$-point functions. We find all odd-point function vanishes due to the parity of the system, i.e.
\begin{equation}
    \langle 
    \widetilde{\delta t}_{a_1} (f_1)
    \cdots \widetilde{\delta t}_{a_{2n+1}}(f_{2n+1}) 
    \rangle
    = 0 
\end{equation}
with $n \in {\mathbb N}^+$. This can be proved inductively. On the other hand, the two point function is given by
\begin{equation}
    \langle\widetilde{\delta t}_a(f_1) 
    \widetilde{\delta t}_b(f_2) 
    \rangle = \delta(f_1 + f_2) \Sigma_{ab}(f_1)\,,
\end{equation}
while the four point function reads
\begin{widetext}
\begin{align}\label{4pt}
    \langle 
    \widetilde{\delta t}_a(f_1) 
    \widetilde{\delta t}_b(f_2) 
    \widetilde{\delta t}_c(f_3) 
    \widetilde{\delta t}_d(f_4) 
    \rangle &= 
    \delta(f_1 + f_2) \delta(f_3 + f_4)
    \Sigma_{ab}(f_1)\Sigma_{cd}(f_3) 
    \\
    &\qquad
    + 
    \delta(f_1 + f_3) \delta(f_2 + f_4)
    \Sigma_{ac}(f_1) \Sigma_{bd}(f_2) 
    \nonumber\\
    &\qquad
    + 
    \delta(f_1 + f_4) \delta(f_2 + f_3)
    \Sigma_{ad}(f_1) \Sigma_{bd}(f_2) 
    \nonumber\\
    &\qquad
    + \delta(f_1 + f_2 + f_3 + f_4) T_{abcd}(f_1,f_2,f_3,f_4)\,,
    \nonumber
\end{align}
where the spectrum, $\Sigma_{ab}$, and the trispectrum, $T_{abcd}$, are given by
\begin{align}
    \Sigma_{ab}(f) &= 
    \frac{1}{(2\pi f)^4}
    \frac{g^2 \bar \rho^2}{\Mpl^4 m^3_\phi} 
    \frac{\nhat_a \cdot \nhat_b}{3} (2 \pi f \tau )^2 K_2(|2\pi f\tau| ) , 
    \\
    T_{abcd}(f_1,f_2,f_3,f_4) 
    &\simeq
    \left[ \prod_{i=1}^4\frac{1}{(2\pi f_i)^2} \right]
    \frac{\pi}{2} \frac{g^4 \bar \rho^4}{\Mpl^8 m^6_\phi}\tau
    \Big[
    (\nhat_a \cdot \nhat_b) (\nhat_c \cdot \nhat_d)
    + (\nhat_a \cdot \nhat_c) (\nhat_b \cdot \nhat_d)
    \Big]. 
    \label{tri_approx}
\end{align}
\end{widetext}
The spectrum is defined as a two-sided spectrum, and hence $\Sigma_{ab}(f) = S_{ab}^{\scriptscriptstyle\rm DM}(f)/2$ with $S_{ab}^{\scriptscriptstyle\rm DM}(f)$ given in Eq.~\eqref{Spectrum_Angular}. The expression for the trispectrum is valid for $2\pi f_i \tau \ll 1$. Here, $\tau = 1 / m_\phi \sigma^2$ is the coherence time scale for the ULDM field. The existence of the non-vanishing trispectrum $T_{abcd}(f_1, f_2,f_3,f_4)$ indicates that the timing residual $\widetilde{\delta t}(f)$ is not Gaussian. 

The achromatic timing residuals are modeled as
$$
\delta t_a(t)
= \sum_{k=1}^{N_f}
\Big[ s_{ak} \sin(2\pi f_k t)
+ c_{ak} \cos(2\pi f_k t)
\Big]
$$
where $f_k = k / T$ and $T$ is the total time baseline of pulsar timing observation. In the main analyses, the coefficients $(s_{ak}, c_{ak})$ are assumed to follow the Gaussian distribution with its covariance matrix given by 
\begin{align}
\langle s_{ak} s_{bk'} \rangle
= \langle c_{ak} c_{bk'} \rangle
&= 
\delta_{kk'} [ 4 \Delta f \Sigma_{ab}(f_k) ] , 
\\
\langle s_{ak} c_{bk'} \rangle 
&= 0 . 
\end{align}

Note that the coefficients $(s_{ak}, b_{ak})$ is nothing but the imaginary and real part of the Fourier component of timing residual. More specifically, one finds
\begin{align}
s_{ak} &= 
2 \Delta f 
\, {\rm Im}[\widetilde{\delta t}_a(f_k)] , 
\\
c_{ak} &= 
2 \Delta f
\, {\rm Re}[\widetilde{\delta t}_a(f_k)] . 
\end{align}
From Eq.~\eqref{4pt}, the four-point function of $s_{ak}$ and $c_{ak}$ contains a non-Gaussian contribution. For instance, we find
\begin{widetext}
\begin{align}
\langle c_{k_1} c_{k_2} c_{k_3} c_{k_4} \rangle
&=
\langle c_{k_1} c_{k_2} \rangle 
\langle c_{k_3} c_{k_4} \rangle
+ 
\langle c_{k_1} c_{k_3} \rangle 
\langle c_{k_2} c_{k_4} \rangle
+
\langle c_{k_1} c_{k_4} \rangle 
\langle c_{k_2} c_{k_3} \rangle
\nonumber\\
&+
2 (\Delta f)^3
\Big[
\delta_{1,234}  T_{abcd}(\bar 1 234)
+ \delta_{2,134} T_{abcd}(1\bar 234)
+ \delta_{3,124} T_{abcd}(12\bar 34)
+ \delta_{4,123} T_{abcd}(123\bar 4)
\Big]
\nonumber\\
&+  2 (\Delta f)^3
\Big[
\delta_{12,34} T_{abcd}(\bar 1 \bar 2 3 4 )
+ \delta_{13,24} T_{abcd}(\bar 1 2 \bar 3 4 )
+ \delta_{14,23} T_{abcd}(\bar 1 2 3 \bar 4 )
\Big]
\end{align}
\end{widetext}
where $T_{abcd}(\bar 1 234) = T_{abcd}(- f_1, f_2,f_3,f_4$) and $\delta_{1,234} = \delta(f_1 - f_2 - f_3 - f_4)$. A similar expression can be obtained for $\langle s_{k_1} s_{k_2} s_{k_3} s_{k_4} \rangle$ and $\langle s_{k_1} s_{k_2} c_{k_3} c_{k_4}\rangle$. Note that 4-pt functions with an odd number of $s_k$ or $c_k$ vanish identically. 

To estimate the degree of non-Gaussianity, we compute the kurtosis of $c_{k}$ for $a = b = c = d$. Note that the skewness vanishes due to the parity symmetry of the system. We find
\begin{align}
{\rm Kurt}(c_k)
= \frac{\langle c_k^4 \rangle}{\langle c_k^2 \rangle^2}
\simeq 3 
\left[ 1 + \frac{9}{64} (2 \pi \Delta f \tau) \right] . 
\end{align}
For the Gaussian random variable, we expect ${\rm Kurt}(X) =3$. The deviation from Gaussianity is thus estimated by the second term in the squared parentheses, i.e. $(9/64) (2\pi \Delta f \tau)$. Since our stochastic signal analysis assumes sampling of at least a few stochastic fluctuations during the total time baseline of pulsar timing observation, $2\pi \Delta f \tau < 1$,  we expect that the deviation from the Gaussianity is sufficiently small. This justifies our treatment of the stochastic signals in the main text. 

\section{Matter Effect}\label{sec:matter_effect_appx}
Quadratic interactions modify the dark matter field profile inside dense astrophysical objects. This occurs because quadratic interactions change the dark matter mass as a function of ambient matter density, therefore altering dark matter field propagation within the objects. As a result, the field profile is modified and ULDM-induced timing signals deviate from those discussed in the main text. We emphasize that the spectra in the main text are obtained under the assumption that astrophysical objects can be treated as point-like particles and that dark matter field propagates like plane wave.

The plane wave assumption is invalid in the regime where the quadratic couplings significantly alter the in-medium scalar mass. In this appendix, we analyze the matter effect in detail, following Refs.~\cite{Day:2023mkb,Banerjee:2025dlo,delCastillo:2025rbr,Gan:2025nlu,Burrage:2025grx}. From this analysis, we derive the form factors that modify the signal power spectra of the timing residual for the Doppler, clock, and pulsar spin signals.

Quadratic interactions generically modify the in-medium mass of dark matter. To illustrate this, consider the quadratic coupling between dark matter and the Standard Model sector as shown in Eq.~(\ref{eq:L_OSM_main}). Inside an astrophysical object, finite-density effects cause dark matter to acquire an in-medium mass,
\begin{align}
m_\tot^2(\boldsymbol x)
= m_\phi^2 + \Delta m^2(\boldsymbol x),  
\end{align}
as shown in Eqs.~\eqref{eq:msq_tot_eq_min}--\eqref{eq:Del_mMSq_main}. As a concrete example, let us consider the ULDM quadratically coupled to the nucleon field. The interaction Lagrangian takes the form as $\mathcal{L} = - g_N m_N \bar{N} N \phi^2/2 \Mpl^2$. In a medium, one can have $\langle \bar{N} N\rangle \simeq n_N$, where $n_N$ is the nucleon number density. In this case, we have $\Delta m^2 = g_N \rho_N/\Mpl^2$.

Ordinary matter affects the evolution of the field, and hence the field profile near such astrophysical objects. The dynamics is governed by the Klein–Gordon equation, 
\begin{align}
\big[ \square + m^2_\tot({\boldsymbol x})\big] \phi(t,\boldsymbol x) = 0 . 
\end{align}
For ULDM, we take the non-relativistic limit. Expanding the field as
\begin{align}
\phi(t, \boldsymbol x)
= \sum_{\boldsymbol k}
\frac{1}{\sqrt{2m_\phi V}} 
\left[
a_{\boldsymbol k} \psi_{\boldsymbol k}(\boldsymbol x) e^{-i \omega_{\boldsymbol k} t}
+ {\rm h.c.}
\right],
\label{field_expansion_matter}
\end{align}
we find that the Klein-Gordon equation reduces to a time-independent Schr{\"o}dinger equation for the mode function $\psi_{\boldsymbol k}(\boldsymbol x)$:
\begin{align}
E_{\boldsymbol k} \psi_{\boldsymbol k}(\boldsymbol x)
= \left[
- \frac{\vecnabla^2}{2m_\phi} + V_\text{eff}(\boldsymbol x) 
\right] \psi_{\boldsymbol k}(\boldsymbol x).
\end{align}
Here, the effective potential is
\begin{align}
V_\text{eff}(\boldsymbol x) = \frac{\Delta m^2(\boldsymbol x)}{2m_\phi}.
\end{align}
In the absence of quadratic interactions, the potential vanishes, and the solutions are simple plane waves. When quadratic interactions are present, the incoming plane wave scatters off the target, leading to a distorted field profile both inside and around the object.
\begin{figure}
    \centering
    \includegraphics[width=0.45\textwidth]{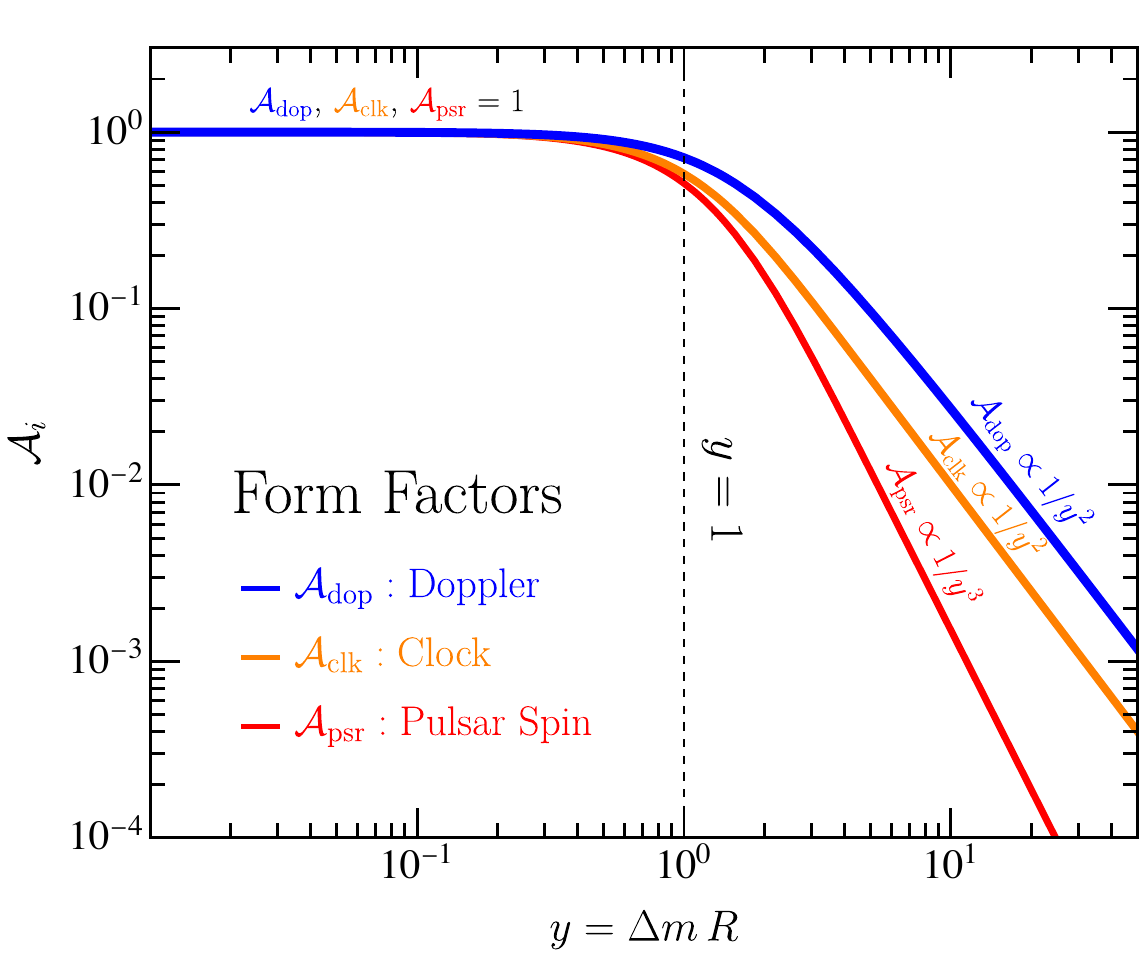}
    \caption{Form factors of the Doppler (blue), clock (orange), and pulsar spin (red) signals for repulsive quadratic interactions, as defined in Eqs.~(\ref{eq:form_factor_dop})--(\ref{eq:form_factor_psr}). Dimensionless parameter $y$ quantifies the strength of the quadratic coupling. For $y \lesssim 1$, all three form factors approach unity, whereas for $y \gtrsim 1$, they exhibit power-law suppression due to the screening effect.}
    \label{fig:form_factors}
\end{figure}

To proceed, we assume that the astrophysical objects we consider --- the Sun, Earth, and pulsars --- are spherically symmetric and uniform in density. Accordingly, we work in spherical coordinates, with ${\boldsymbol x} = r \, \rhat$, where $\rhat$ is the radial unit vector. In this approximation, the effective potential takes the form
\bea
V_\text{eff}(r) = V_0 \, \Theta(R -r).
\eea
The wave function is written as
\begin{equation}
\psi_{\boldsymbol k}(\boldsymbol x) = 
\left\{
\begin{aligned}
& \psi^{\rm out}_{\boldsymbol k}(\boldsymbol x) & r \geq R \\
& \psi^{\rm in}_{\boldsymbol k}(\boldsymbol x) & r \leq R
\end{aligned}
\right. ,
\end{equation}
where $\psi_{\boldsymbol k}^{\rm out}(\boldsymbol x)$ and $\psi_{\boldsymbol k}^{\rm in}(\boldsymbol x)$ denote the solutions outside and inside the dense object, respectively. In a spherically symmetric system, the wave function is generally expressed as
\begin{align}
\label{eq:psi_spatial_appx}
\psi_{\boldsymbol k}(\boldsymbol x) = 
R_\ell(r,k)\,P_{\ell}(\khat \cdot \rhat) ,
\end{align}
where $R_\ell(r,k)$ is the radial wavefunction and $P_\ell(\khat \cdot \rhat)$ is a Legendre polynomial. The radial Schr{\"o}dinger equation takes the form
\bea
\left[
\frac{d^2}{dr^2}
+ \frac{2}{r} \frac{d}{dr}
+ k^2 - \frac{\ell(\ell+1)}{r^2}
\right] R_\ell^{\rm out}(kr) = 0
\eea
for the wavefunction outside the sphere, and 
\bea
\left[
\frac{d^2}{dr^2}
+ \frac{2}{r} \frac{d}{dr}
+ p^2 - \frac{\ell(\ell+1)}{r^2}
\right] R_\ell^{\rm in}(pr) = 0
\eea
for the wavefunction inside the sphere. In the equation above,
\begin{align}
\label{eq:p_def_appx}
p = \sqrt{k^2 - \Delta m^2}
\end{align}
denotes the momentum inside the sphere. A general solution to the radial equation can be written as
$$
R_\ell^{{\rm out},\,{\rm in}}(x) = c_1 j_\ell(x) + c_2 y_\ell(x) = d_1 h^{(1)}_\ell(x) + d_2 h^{(2)}_\ell(x),
$$
where $j_\ell(x)$ and $y_{\ell}(x)$ are spherical Bessel functions, and $h_\ell^{(1,2)}(x) = j_\ell(x) \pm i y_\ell(x)$ are the spherical Hankel functions of the first and second kind. The wave functions outside and inside the sphere are given by
\begin{align}
\label{eq:partial_wave_out}
\psi_{\boldsymbol k}^{\rm out}(\boldsymbol x)
&= \sum_{\ell=0}^\infty (2\ell +1 ) \, i^\ell \,
\big[ j_\ell(kr) \\
& \quad \quad \quad \quad \,+ c_\ell^{\rm out}(k) \, h_\ell^{(1)}(kr) \big] \, P_\ell(\khat \cdot \rhat), \nonumber\\
\psi^{\rm in}_{\boldsymbol k}(\boldsymbol x)
& = 
\sum_{\ell=0}^\infty ( 2\ell +1) \, i^\ell \, c^{\rm in}_{\ell}(k) \, j_\ell(p r) \, 
P_\ell(\khat \cdot \rhat). 
\label{eq:partial_wave_in}
\end{align}
In the exterior solution, the first term in parentheses describes the incident plane wave $e^{i \boldsymbol k \cdot \boldsymbol x}$, while the second term represents the outgoing scattered wave, which can be seen from its asymptotic behavior $\lim_{r\to\infty}h_\ell^{(1)}(kr) \propto e^{ikr}/kr$. In the interior solution, we keep only $j_\ell(x)$ since $y_\ell(x)$ diverges at $r=0$ and is therefore unphysical.

The coefficients $c_\ell^{\rm out}$ and $c_{\ell}^{\rm in}$ are obtained by matching the exterior and interior wave functions at the boundary. The continuity conditions are
\begin{align}
\psi^{\rm out}_{\boldsymbol k}(R) = \psi^{\rm in}_{\boldsymbol k}(R) ,  \quad \frac{d\psi^{\rm out}_{\boldsymbol k}}{dr} (R) = 
\frac{d\psi^{\rm in}_{\boldsymbol k}}{dr} (R). 
\end{align}
These give
\begin{align}
\!\!\! 
c_{\ell}^{\rm out} (k)
&= 
- \frac{k j_\ell(pR) \, j_{\ell+1}(kR) - p j_{\ell}(kR) \, j_{\ell+1}(pR)}{k j_\ell(pR) \, h^{(1)}_{\ell+1}(kR) - p h^{(1)}_{\ell}(kR) \, j_{\ell+1}(pR)},
\\
\!\!\! 
c_\ell^{\rm in} (k)
&=
- \frac{ik \, (kR)^{-2}}{ k  j_\ell(pR) h^{(1)}_{\ell+1}(kR) - p h^{(1)}_{\ell}(kR) \, j_{\ell+1}(pR)},
\end{align}
which agree with previous literature, e.g., Refs.~\cite{Day:2023mkb,Banerjee:2025dlo,delCastillo:2025rbr,Gan:2025nlu,Burrage:2025grx}. In the absence of a potential~($\Delta m = 0$), the exterior and interior wavenumbers coincide, $k=p$. The coefficients then reduce to $c^{\rm in}_\ell=1$ and $c^{\rm out}_\ell=0$, which reproduces the plane-wave solution $e^{i \boldsymbol k \cdot \boldsymbol x}$.

\subsection{Doppler Signal}

In this subsection, we discuss the matter effect on the Doppler signal, which arises from the perturbation of the Solar System barycenter and is dominated by fluctuations in the position of the Sun.

We begin by considering a generic astrophysical object, which experiences a ULDM-induced force per unit mass given by
\begin{align}
\label{eq:force_eq_ave}
\bm{\mathcal{F}} = - \frac{1}{\volR}\int_{\volR} d^3 {\boldsymbol x} \, \frac{\vecnabla \rho}{\rho}.
\end{align}
Here, $\rho$ denotes the density, and $\volR = 4 \pi R^3/3$ denotes the volume of the astrophysical object, respectively. Unlike Eq.~(\ref{force_eq}) in the main text, here we account for the distortion of the field configuration inside the object from a plane wave due to the matter effect.

Recalling that the relative density variation is $\delta\rho/\rho=g\,\varphi^2$, and using~\eqref{field_expansion_matter}, we obtain
\begin{align}
\pmb{\cal F}
= - \frac{g}{2\Mpl^2 \mphi V}
\sum_{\veck,\veck'} a_{\veck} a_{\veck'}^* 
e^{-i (\omega_{\boldsymbol k} - \omega_{\boldsymbol k'}) t} 
\bm{H}(\veck, \veck').
\end{align}
Here we kept only the slow mode fluctuations. In the above equation, the vector $\bm{H}(\veck, \veck')$ depends on the incident momenta $\veck$ and $\veck'$, and is defined as
\begin{align}
\label{eq:H_kkp_1}
\bm{H}(\veck, \veck')
=
\frac{1}{\volR}
\int_{\volR} d^3 {\boldsymbol x} \, 
\vecnabla \big[ 
\psi^{\rm in}_{\veck}(\boldsymbol x)
\psi^{\rm in \, *}_{\veck'}(\boldsymbol x)
\big].
\end{align}
Neglecting the pulsar term, which is strongly suppressed by screening inside the pulsar, we obtain the timing-residual power spectrum,
\begin{widetext}
\begin{align}
S_{ab}^{\rm DM}(f)
= \frac{1}{(2\pi f)^4}
\frac{g^2}{2\Mpl^4 m^2_\phi}
\int d^3 {\boldsymbol v'} \, d^3 {\boldsymbol v''}
f(\bm{v}')
\,f(\bm{v}'')
\,[\nhat_a \cdot \bm{H}(\veck', \veck'')]
\,[\nhat_b \cdot \bm{H}^*(\veck', \veck'')]
\,\delta(f-f_{\boldsymbol k'}+f_{\boldsymbol k''}).
\end{align}
\end{widetext}

The power-spectrum computation above involves $\bm{H}(\veck,\veck')$, which is defined by a volume integral over the gradient of partial-wave products. Expanding $\psi^{\rm in}_{\veck}(\boldsymbol x)$ to the s-wave~($l=0$) and p-wave~($l=1$) based on Eq.~\eqref{eq:partial_wave_in}, we have
\begin{equation}
\label{eq:psi_s_p_exp_appx}
\psi^{\rm in}_{\veck}(\boldsymbol x) \simeq \widetilde{\psi}^\text{in}_{0}(k,r) + \widetilde{\psi}^\text{in}_{1}(k,r) 
(\khat \cdot \rhat),
\end{equation}
where $\widetilde{\psi}^\text{in}_{0}(k)$ and $\widetilde{\psi}^\text{in}_{1}(k)$ are the radial components of the $l=0$ and $l=1$ partial waves. From Eq.~(\ref{eq:partial_wave_in}), we have
\bea
\label{eq:l_01_radial_appx}
&\widetilde{\psi}^\text{in}_{0}(k,r) = c_0^\text{in}(k) \, j_0(pr),\\
&\widetilde{\psi}^\text{in}_{1}(k,r) = 3 i \, c_1^\text{in}(k) \, j_1(pr) .
\eea
Higher partial waves are neglected since the system we consider is deeply in the low-energy regime, $kR \ll 1$, and $|\widetilde{\psi}_{l+1}^{\rm in}|/|\widetilde{\psi}_{l}^{\rm in}| \sim \mathcal{O}(kR)$. As the Doppler signal involves a gradient of the field, the pure $s$-wave component cancels, and thus, we focus on the mixing between the $s$- and $p$-wave components, which provides the non-vanishing leading contribution to $\bm{H}(\veck,\veck')$. 

Including the $\ell=0$ and $\ell=1$ partial waves, we find
\begin{align}
\label{eq:Hkkp_Ad_1}
\boldsymbol H(\boldsymbol k, \boldsymbol k')
= {\cal A}_\dop \times \, i (\boldsymbol k - \boldsymbol k').
\end{align}
Substituting Eq.~\eqref{eq:psi_s_p_exp_appx} into Eq.~\eqref{eq:H_kkp_1}, and using the identity
\bea
& \,\,\,\,\,\, \frac{1}{\mathcal{V}_R} \int_{\mathcal{V}_R} d^3 {\boldsymbol x} \, \vecnabla[F(r) \, (\khat \cdot \rhat)] \\
& = \frac{1}{R^3} \int^R_0 dr \frac{d [r^2 F(r)]}{dr} \, \khat,
\eea
where $F(r)$ is an arbitrary function of $r$, we obtain the form factor of the Doppler signal
\begin{equation}
\label{eq:Ad_int_appx}
{\cal A}_\dop = \frac{1}{i k R^3} \int_0^R dr \frac{d}{dr}\left[ r^2 \, \widetilde{\psi}^\text{in}_{0}(k',r) \, \widetilde{\psi}^\text{in}_{1}(k,r) \right].
\end{equation}
Taking the $kR, k'R \ll 1$ limit, we obtain
\begin{align}
{\cal A}_\dop(y)
\simeq 3 \frac{y- \tanh y}{y^3}.
\end{align}
The resulting Doppler-term timing-residual power spectrum is
\begin{align}
S_{ab}^{\rm DM} (f) 
= [{\cal A}_\dop(y)]^2
\, [S_{ab}^{\rm DM}(f)]_0 , 
\end{align}
where $[S_{ab}^{\rm DM}(f)]_0$ denotes the signal power spectrum in the absence of the matter effect.

When $y \lesssim 1$, the matter effect is negligible. In this regime, we have ${\cal A}_\dop(y)\simeq 1$. This follows from the fact that the wavefunction is approximately a plane wave, $\psi_{\boldsymbol k}(\boldsymbol x) \simeq e^{i \boldsymbol k \cdot \boldsymbol x}$. Substituting this into Eq.~(\ref{eq:H_kkp_1}), we obtain $\bm{H}(\veck, \veck') = i (\veck - \veck')$, reproducing the standard result quoted in the main text.

When $y \gtrsim 1$, the matter effect becomes significant. For the Doppler signal, the dominant contribution comes from the fluctuations of solar system barycenter. In this case, the condition for the matter effect to be apparent is
\begin{align}
\label{eq:matter_line_appx}
| g_\odot | 
& \gtrsim 4\times 10^5 \times
\bigg( \frac{1\,{\rm g/cm^3}}{\rho_\odot} \bigg).
\end{align}
Above this critical value, the matter effect suppresses the signal power spectrum, and the Doppler signal therefore saturates to a constant, independent of the coupling strength.

\subsection{Clock Signal}

Since Terrestrial Time is defined by an ensemble of atomic clocks operating on the Earth's surface, the clock signal is inevitably affected by the matter effect. This effect depends on the scalar-field value at the Earth's surface. In the following, we evaluate the matter effect on the coherent clock signal.

The timing residual induced by the fluctuations of Terrestrial Time can be similarly parametrized as
\begin{align}
\delta t_a(t) = \frac{g_{\scriptscriptstyle\rm TT}}{2 \Mpl^2} \int^t dt' \, \phi^2(t').
\nonumber 
\end{align}
To examine how the clock signals are affected by the density of Earth, we recall the mode expansion evaluated on the Earth surface:
\begin{align}
\label{eq:phi_earth_clock_appx}
\phi(t) = \frac{1}{\sqrt{V}}\sum_{\boldsymbol k} a_{\boldsymbol k} \psi_{\boldsymbol k}(R_\oplus) \, e^{-i m_\phi t} + {\rm h.c.}.
\end{align}
Since we are deeply in the low-energy limit $kR_\oplus \ll1$, we may approximate the wave function with the leading s-wave component, $\psi_{\boldsymbol k}(R_\oplus) \simeq  \widetilde{\psi}_{0}(R_\oplus)$. With this approximation, we can factor out $\psi_{\boldsymbol k}(R_\oplus)$ and express Eq.~\eqref{eq:phi_earth_clock_appx} as
\bea
\phi(t) \simeq \widetilde{\psi}_{0}(R_\oplus) \times {\sqrt{\frac{2}{m_\phi}}}{\cal R} \cos(m_\phi t - \Theta),
\eea
where ${\cal R}$ and $\Theta$ are defined in Eq.~\eqref{RTheta}. 
This leads to the following timing residual:
\begin{equation}
\label{eq:delta_t_clock_appx}
\delta t_a(t) \simeq [\widetilde{\psi}_{0}(R_\oplus)]^2 \times \frac{ g_{\scriptscriptstyle\rm TT}}{4 \Mpl^2 m_\phi^{2}} {\cal R}^2 \sin(2 m_\phi t - 2 \Theta).
\end{equation}

Based on Eq.~\eqref{eq:delta_t_clock_appx}, the clock signal can be written as
\begin{align}
\delta t_a(t) = {\cal A}_\clk(y) \times [\delta t_a(t)]_0,
\end{align}
where the form factor is given by
\begin{align}
{\cal A}_\clk(y)
\simeq [\widetilde{\psi}_{0}(R_\oplus)]^2
\simeq \frac{\tanh^2 y}{y^2}. 
\end{align}
Note that $[\delta t_a(t)]_0$ is the timing residual without the matter effect. The power spectrum is given by
\bea
S^{\scriptscriptstyle\rm DM}_{ab}(f) = [{\cal A}_\clk(y)]^2 [S^{\scriptscriptstyle\rm DM}_{ab}(f)]_0.
\eea
The matter effect for the clock signal becomes important for
\begin{align}
| g_\oplus | 
& \gtrsim 10^9 \times
\bigg( \frac{5\,{\rm g/cm^3}}{\rho_\oplus} \bigg) . 
\end{align}
Since pulsar timing can probe dilaton couplings much smaller than the above critical values relevant for the coherent-signal search, the matter effect is negligible across most of the parameter space of interest.

\subsection{Pulsar Spin Signal}

The pulsar spin signal arises from fluctuations in the pulsar mass and its radius. The fluctuation in its total mass can be estimated as
\begin{align}
\frac{\delta M}{M} =
\, \frac{ \sum_i \int d^3 {\boldsymbol x} \, \delta \rho_i(\boldsymbol x)}{\sum_i \int d^3 {\boldsymbol x} \, \rho_i(\boldsymbol x)}.
\end{align}
where $i=n,p,\mu,e$, and $\rho_i(\boldsymbol x)$ is the number density of each constituent. Assuming that the density is uniform within the sphere, we find
\begin{equation}
\frac{\delta M}{M} = \sum_i \mathcal{X}_{\psr,i} \times \frac{1}{\volR} \int_{\volR} d^3 {\boldsymbol x} \, \frac{\delta \rho_i(\boldsymbol x)}{\rho_i},
\end{equation}
where $\mathcal{X}_{\rm psr,i}$ is the mass fraction of the constituent $i$. The volume average of the mass density fluctuation induced by ULDM is then given by
\begin{align}
&\quad \,\,\, \frac{1}{\volR} \int_{\volR} d^3 {\boldsymbol x} \, \frac{\delta \rho_i(\boldsymbol x)}{\rho_i}
\nonumber\\
& = \frac{g_{i}}{4\Mpl^2 m_\phi V}
\sum_{\boldsymbol k, \boldsymbol k'}
a_{\boldsymbol k} 
a_{\boldsymbol k'}
\, e^{-2im_\phi t} \, 
{\cal A}_\psr
+ {\rm h.c.},
\end{align}
where $\volR$ is the volume of the pulsar, and the corresponding form factor is
\begin{align}
{\cal A}_\psr
= \frac{1}{\volR}
\int_{\volR} d^3 {\boldsymbol x} \, \psi^{\rm in}_{\boldsymbol k}(\boldsymbol x) \psi^{\rm in}_{\boldsymbol k'}(\boldsymbol x).
\end{align}
As for the clock signal, we only keep the coherent pulsar spin signal. Since we are in the low-energy limit, we again approximate the wave function with the leading s-wave component. With this approximation, we find the form factor of the pulsar mass fluctuation as
\begin{equation}\label{formfactor_pulsar_spin}
{\cal A}_\psr(y) \simeq \frac{3}{2} \left(\frac{\tanh^2 y}{y^2} - \frac{y - \tanh y}{y^3}\right),
\end{equation}
where $y$ is defined in Eq.~(\ref{eq:y_parameter}).

The pulsar spin signal also arises from the pulsar radius fluctuation, as indicated in Eq.~\eqref{eq:z_M_R_mn}. The pulsar radius is determined by the balance between the degenerate pressure of neutron and the gravitational force. The degenerate pressure depends on the mass of neutron in the bulk. To compute the response of the radius with respect to the propagating dark matter, we replace the neutron mass variation with the volume-averaged one. Therefore, the form factor for the pulsar radius fluctuation is identical to Eq.~\eqref{formfactor_pulsar_spin}. Consequently, the form factor for the pulsar spin fluctuation is Eq.~\eqref{formfactor_pulsar_spin}, reproducing the result listed in Eqs.~\eqref{eq:form_factor_dop}--\eqref{eq:form_factor_psr} of the main text.

The matter effect for the pulsar spin fluctuation becomes important when
\begin{equation}
    |g_{\rm psr}| \gtrsim 5\times
    \bigg( \frac{ \lambdaqcd^4 }{\rho_\text{psr}} \bigg)
\end{equation}
where we use $R_\text{psr}=10\,{\rm km}$ and $\lambdaqcd = 200\MeV$ for pulsars. For the low-mass end of the coherent signal search, the matter effect seems irrelevant, while for the higher-mass end, it might become important.

\subsection{Light QCD Axion}

For light QCD axion models, the quadratic interaction leads to an attractive potential for the propagation of the dark matter field. From Eq.~\eqref{axion_quad}, the in-medium mass squared can be obtained as
\begin{align}
m_\tot^2(\boldsymbol x)
= m_\phi^2+ \frac{C_N \rho_N(\boldsymbol x)}{\fphi^2},
\label{axion_inmedium_mass}
\end{align}
where ${C_N \sim - 10^{-2}}$~\cite{Ubaldi:2008nf,Kim:2022ype}. When the Klein-Gordon equation for the axion is expressed in the Schr{\"o}dinger form as Eq.~\eqref{eq:schrodinger_matter}, the above in-medium mass squared acts as an attractive potential due to the negative sign of $C_N$.

The above discussion assumes that the minimum of the axion potential inside the matter is the one that conserves CP symmetry in the strong sector, i.e. $\phi = 0$. When the finite density correction is large, such correction to the axion potential would trigger the phase transition inside matter~\cite{Hook:2017psm}. To see how this changes the attractive nature of the potential, we consider a simple phenomenological model of light QCD axions, discussed in Ref.~\cite{Hook:2017psm}. The bare potential is
\begin{align}
V_0(\phi) = -\eps m_\pi^2 f_\pi^2
\sqrt{1 - \frac{4 m_u m_d}{(m_u + m_d)^2} \sin^2\left(\frac{\phi}{2\fphi}\right)},
\label{zero_temp_potential}
\end{align}
where $m_\pi$ and $f_\pi$ are pion mass and decay constant. Here, we consider the light QCD axion regime, i.e. $\eps \ll1$; when $\eps=1$, the above reproduce a standard QCD axion potential. In the limit $m_u = m_d$, the axion mass is $m_\phi^2 \fphi^2 = \eps m_\pi^2 f_\pi^2 / 4$.
\begin{figure}
\centering
\includegraphics[width=0.43\textwidth]{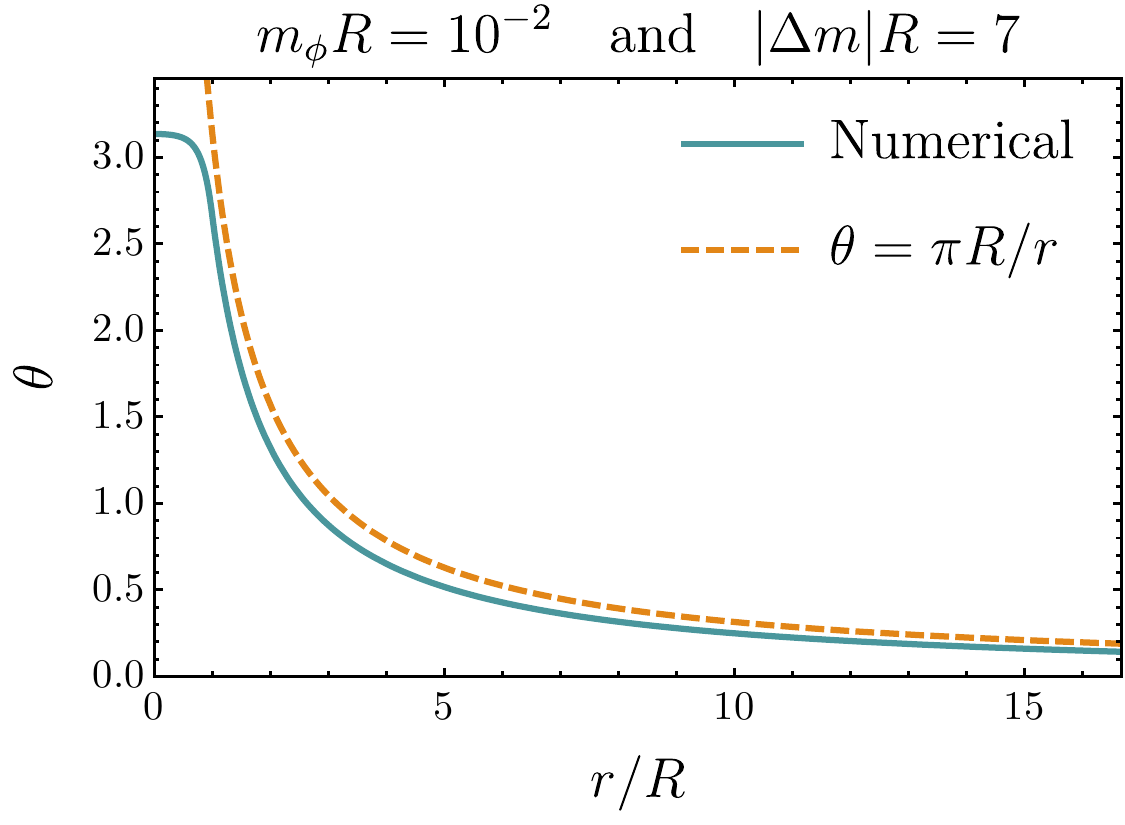}
\caption{Static QCD axion profile sourced by a dense object. We choose the $|\Delta m| \, R = 7$ and $m_\phi R = 10^{-2}$ for this plot, where $\Delta m$ is the finite density correction to the axion mass squared evaluated at $\theta_0 = \pi$. Inside  matter $r/R <1$, the profile approaches $\theta_0 = \pi$, while outside matter $r/R>1$, it behaves as $1/r$.}
\label{fig:static_profile}
\end{figure}

Inside ordinary matter, the axion potential acquires a finite-density correction with the opposite sign to the bare potential. This correction is given by
\begin{align}
V_N(\phi) = \sigma_{\pi N} n_N
\sqrt{1 - \frac{4 m_u m_d}{(m_u + m_d)^2} \sin^2\left(\frac{\phi}{2\fphi}\right) },
\label{finite_density_correction}
\end{align}
where $\sigma_{\pi N} \simeq \partial \, m_N/ \partial \ln m_\pi^2 \sim 50\,\MeV$ is a nucleon $\sigma$-term~\cite{FlavourLatticeAveragingGroupFLAG:2024oxs}, and $n_N$ is the nucleon number density. The above finite density correction may be understood from the $\theta$-dependent nucleon mass, ${\cal L} = - m_N(\theta) \bar NN$. By expanding the nucleon mass term around $\theta=0$, one finds
\begin{align}
{\cal L} = - m_N(\theta) \bar N N
&\supset - \sigma_{\pi N} \frac{\delta m_\pi^2(\theta)}{m_\pi^2(0)} \bar NN,
\label{quadratic_axion}
\end{align}
which leads to the finite density correction in Eq.~\eqref{finite_density_correction}.

As discussed in Ref.~\cite{Hook:2017psm}, the axion undergoes a phase transition once the potential energy gain overcomes the field-gradient energy characterized by its bare mass, and the finite-density correction exceeds the vacuum contribution, i.e., $|\Delta m| > R^{-1}$ and $|\Delta m| > m_\phi$. These two conditions can be written in terms of the axion mass and its decay constant,
\begin{align}
\frac{1}{\fphi^2} \gtrsim
\frac{1}{\sigma_{\pi N} n_N} 
\max(4 m_\phi^2 , \, 1/R^2)
\end{align}
where $R$ is the size of the object under consideration. When this condition is met, the phase transition is triggered and a static axion profile develops as
\begin{align}
\label{eq:axion_static_profile}
\theta_0(r) \simeq \pi 
\begin{cases}
1 & r \leq R
\\
e^{-m_\phi(r-R)} \frac{R}{r} 
& r> R
\end{cases}.
\end{align}
That is, inside the matter, the axion finds a new minimum around $\theta \simeq \pi$ and the field profile outside matter is given by a Yukawa-type potential. A numerical solution of the static profile can be found in Fig.~\ref{fig:static_profile}. The range of parameter space that exhibits the phase transition is shown as a blue region in Fig.~\ref{fig:MICROSCOPE_QCDAxion}.

\begin{figure}
\centering
\includegraphics[width=0.45\textwidth]{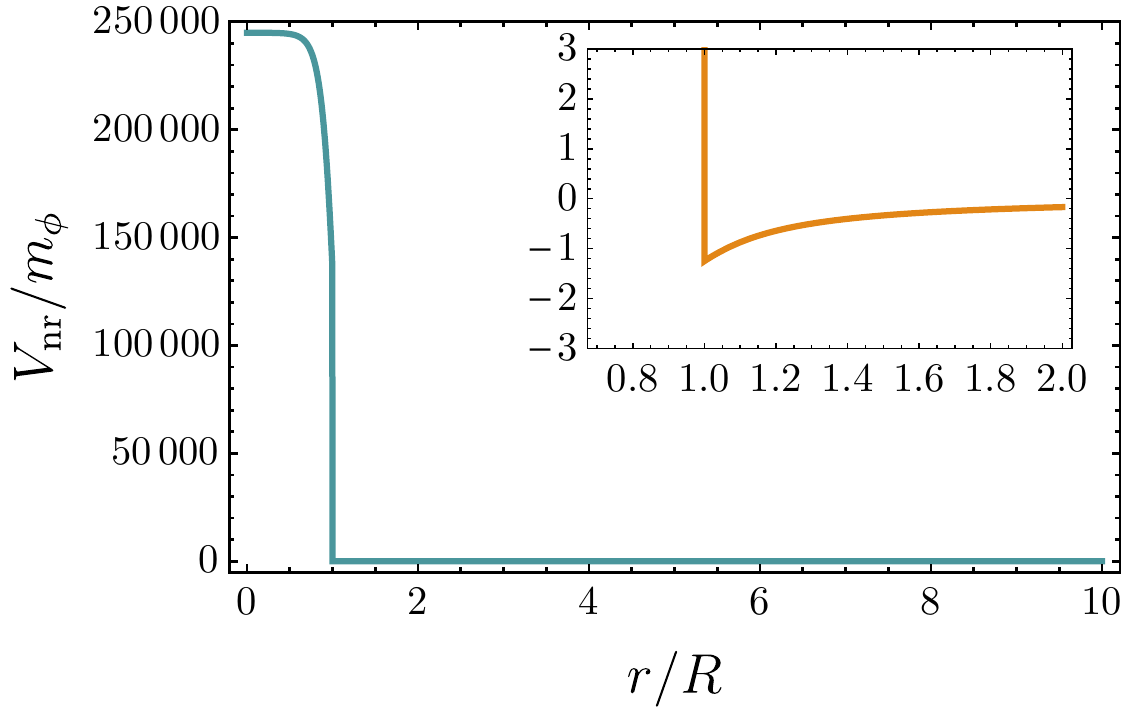}
\caption{An example of $V_{\rm nr}$ when the phase transition occurs. As discussed in the text, due to non-vanishing static profile, the in-medium mass squared takes a positive value, and hence, it acts as a repulsive potential rather than attractive one. Outside potential, we find $-1/r^2$ behavior, which can be neglected for the study of dark matter propagation in the parameter space of interest. For this result, we use the static profile obtained in the previous figure.}
\label{fig:nr_potential}
\end{figure}

This alters the propagation of the dark matter field. Note that the in-medium mass squared \eqref{axion_inmedium_mass} is obtained by expanding $\theta$-dependent nucleon mass \eqref{quadratic_axion} around $\phi=0$. To examine the propagation of the dark matter field in the presence of phase transition, we expand the field around a new static profile as
$$
\phi = \phi_0 + \delta \phi,
$$
where $\phi_0 = \theta_0 f_\phi$ and $\delta\phi$ is fluctuations around it. The equation of motion for dark matter $\delta\phi$ is then given by
\begin{align}
[ \square + V''(\theta_0) ] \delta\phi = 0,
\end{align}
where $V(\theta_0)$ is the potential including the zero-temperature one \eqref{zero_temp_potential} and finite density correction \eqref{finite_density_correction} evaluated at the new static profile $\theta_0$. In the Schr{\"o}dinger form, the non-relativistic potential can be obtained as
\begin{align}
V_{\rm nr} = \frac{V''(\theta_0) - m_\phi^2}{2m_\phi}. 
\end{align}
A numerical example of $V_{\rm nr}$ with numerically obtained static profile $\theta_0$ is given in Fig.~\ref{fig:nr_potential}.

Inside matter, we find
\begin{align}
V''(\pi) = \frac{\sigma_{\pi N} n_N - \eps m_\pi^2 f_\pi^2}{\fphi^2} \frac{m_u m_d}{m_d^2 - m_u^2}. 
\end{align}
Since we are interested in the regime where the phase transition takes place, the above quantity is positive. That is, the in-medium mass squared is positive and the non-relativistic potential behaves as a repulsive potential. 

Outside matter, the potential becomes more complicated. Considering sufficiently large $r$ such that $\theta_0 \ll1$, we may expand the potential around $\theta_0 \sim 0$. We find
\begin{align}
V''(\theta_0) \simeq m_\phi^2
\left[
1 - \frac{\theta_0^2}{2} \frac{m_u^2 + m_d^2 - m_u m_d}{(m_u + m_d)^2} 
\right]. 
\end{align}
The second quantity provides $-e^{-2m_\phi r}/r^2$ potential for the propagation of dark matter outside of the finite density object.

We argue that this $1/r^2$ potential piece can be neglected in our case. To illustrate this, we expand the field $\delta \phi$ as in Eq.~\eqref{field_expansion_matter}. The Klein-Gordon equation becomes
\begin{align}
E_{\boldsymbol k} \psi_{\boldsymbol k}(\boldsymbol x)
= \left[
- \frac{\vecnabla^2}{2m_\phi} 
- \frac{\beta}{m_\phi r^2} e^{-2m_\phi(r-R)}
\right] \psi_{\boldsymbol k}(\boldsymbol x)
\end{align}
where $E_{\boldsymbol k} = |{\boldsymbol k}|^2/2m_\phi$ and 
\begin{align}
\beta = \frac{(m_\phi R)^2}{4}
\frac{m_u^2 + m_d^2 - m_u m_d}{(m_u+m_d)^2}. 
\end{align}
To proceed, we will ignore the exponential factor. By showing that the above potential without the exponential factor is negligible for the parameter space of interest, we will justify that the non-relativistic potential for the light QCD axions when the phase transition occurs can be simply approximated as a finite repulsive potential well. 

Following the usual procedure, we expand the wave function into the radial wave function and spherical harmonics \eqref{eq:psi_spatial_appx}. From the Schr{\"o}dinger equation, we find the radial equation as
\begin{align}
\left[ 
\partial_r^2 + k^2 + \frac{\beta}{r^2} - \frac{\ell(\ell+1)}{r^2} \right] u_{k\ell}(kr) = 0,
\end{align}
where $u_{k\ell}(r) =r R_{k \ell}(r)$. The solution is given by the Bessel functions
\bea
u_{k\ell}(kr) 
& = c_1 \sqrt{kr} J_{\frac12\sqrt{(2\ell+1)^2 -4\beta}}(kr)\\
& \,\, + c_2 \sqrt{kr} Y_{\frac12\sqrt{(2\ell+1)^2 -4\beta}}(kr).
\eea
For all $\ell$, as long as $\beta \ll1$, the solution is approximately given by the radial equation without $1/r^2$ potential piece.

We find that $\beta \ll1$ almost for the entirely mass range of dark matter that we discuss in this work. For the coherent signal search, this is trivially satisfied, as one can see $m_\phi R \ll 1$ for the Earth and pulsar radius $R$. For the stochastic signal search, we find
\begin{align}
\beta \simeq 0.2 \, \Big( \frac{m_\phi}{10^{-16}\eV} \Big)^2 \Big( \frac{R}{R_\odot} \Big)^2. 
\end{align}
As long as the axion bare mass is smaller than $10^{-16}\eV$, one can ignore the potential due to the extended static axion profile for the evolution of dark matter field.

\subsection{MICROSCOPE: Light QCD Axion}\label{subsec:MICROSCOPE_Axion}

In this subsection, we discuss how the matter effect, especially the phase transition inside the Earth, induces an extra axion force, leading to distinctive constraints from equivalence principle violation tests such as MICROSCOPE satellite mission. This satellite is operated at altitude $r - R_\oplus = 710 \, \text{km}$, where $r$ is the distance from the satellite to the Earth's center and $R_\oplus$ is the radius of the Earth. Constraints on dilaton-like scalars have already been presented in Refs.~\cite{Hees:2018fpg, Banerjee:2022sqg,Gan:2025nlu,Gue:2025nxq,Delaunay:2025pho}. Here, we update the corresponding constraint for light QCD axion dark matter, which exhibits qualitatively different behavior. Before we proceed, we recall that the violation of the equivalence principle is parameterized by the E\"otv\"os parameter,
\bea
\eta = \frac{|{\bf a}_\text{Pt}-{\bf a}_\text{Ti}|}{|{\bf g}|},
\eea
where ${\bf a}_\text{Pt}$ and ${\bf a}_\text{Ti}$ are the accelerations of the test masses, composed of platinum and titanium alloys, and ${\bf g} = - G_N M_\oplus \, \hat{{\bf r}}/r^2$ is the local gravitational acceleration. Here, $M_\oplus \simeq 5.97 \times 10^{24} \, \text{kg}$ is the Earth’s mass. According to Ref.~\cite{MICROSCOPE:2022doy}, the reported $1\sigma$ uncertainty is $\sigma_\eta \simeq 2.7 \times 10^{-15}$. In our discussion, we adopt $\eta \lesssim 2 \sigma_\eta$ as an approximate $95 \, \% \, {\rm C.L.}$ constraint.
\begin{figure}
\centering
\includegraphics[width=0.45\textwidth]{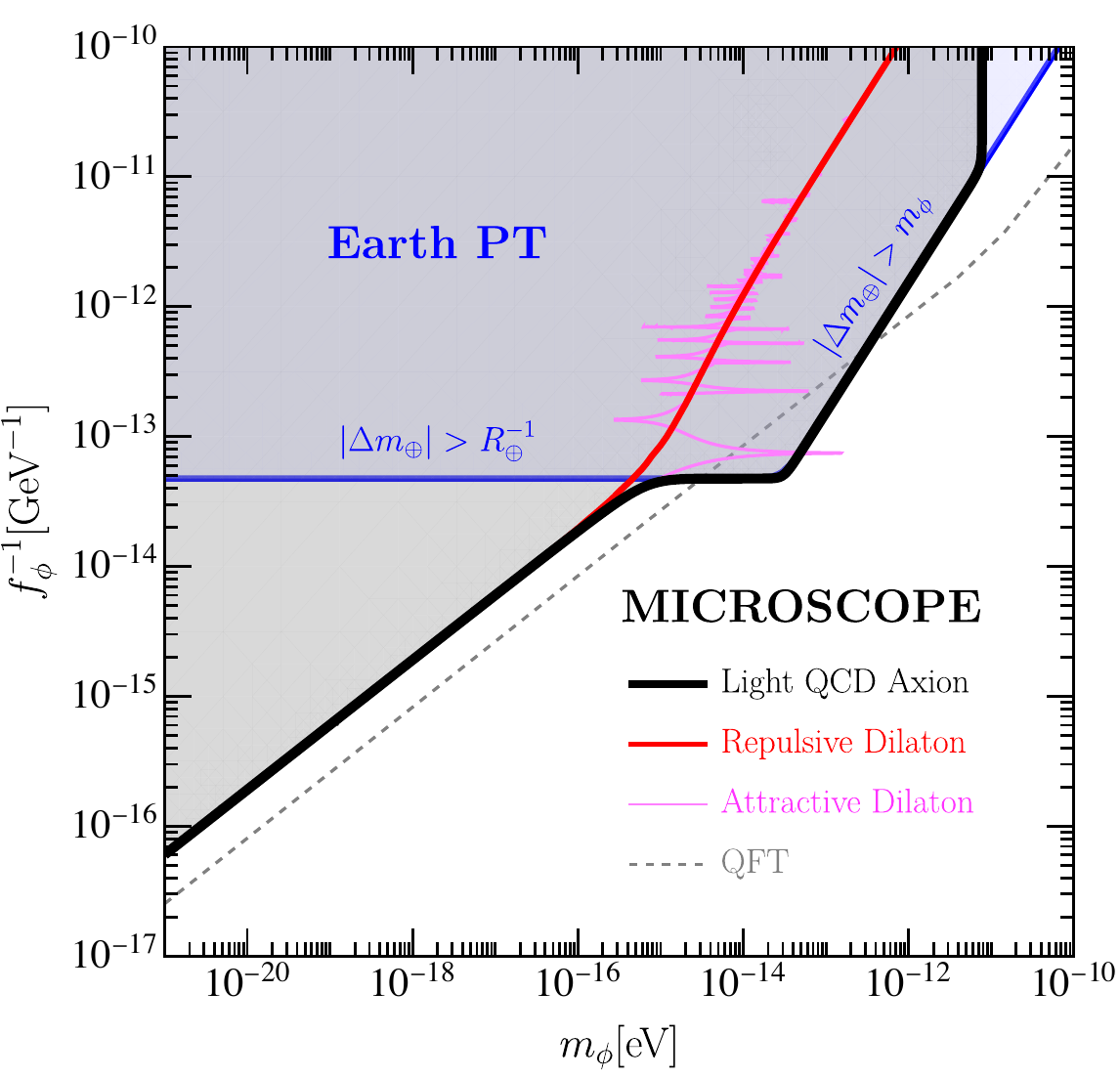}
\caption{The updated MICROSCOPE constraint on the light QCD axion is shown by the gray shaded region enclosed by the thick black line. The red solid line represents the recast MICROSCOPE constraint assuming that $\phi$ is a dilaton with repulsive interactions~\cite{Hees:2018fpg, Banerjee:2022sqg,Gan:2025nlu}, while the magenta solid line represents the same limit assuming attractive interactions~\cite{Gue:2025nxq,Delaunay:2025pho}. The gray dashed line represents the MICROSCOPE constraint obtained from perturbative QFT computations~\cite{Grossman:2025cov}. The region enclosed by the blue solid line indicates the parameter space where a phase transition occurs inside the Earth, as discussed in Ref.~\cite{Hook:2017psm}.}
\label{fig:MICROSCOPE_QCDAxion}
\end{figure}

In the presence of an axion field, the mass of a test body can be expanded as
\begin{align}
\label{eq:M_test_axion_expansion_appx}
M_\test(\theta) = M_\test \left( 1  + \frac{Q_{\test}^\axion}{2} \theta^2 + \cdots \right).
\end{align}
Here, 
\begin{equation}
Q_{\test}^\axion = \frac{\partial \ln M_\test}{\partial \, \theta^2}
\end{equation}
is defined as the axion charge, and  $\test=\text{Pt},\text{Ti}$ labels the test masses. The force between the Earth and a test mass is then
\begin{align}
\label{force_axion_general}
\mathbf{F}_{\test} 
= - \vecnabla M_\test(\theta) 
= - \frac{Q_{\test}^\axion M_\test}{2} \vecnabla \theta^2. 
\end{align}
The axion charge defined above has two contributions: (i) variation of nucleon rest masses induced by $\phi$; (ii) variation of nuclear binding energy induced by $\phi$. The axion charges associated with free nucleon rest masses are taken from Ref.~\cite{Kumamoto:2024wjd}. The axion charges arising from nuclear binding energy are taken from Refs.~\cite{Gue:2024onx,Bauer:2024hfv,Gue:2025nxq} with slightly different conventions. In our convention, they are obtained by making the replacement $Q_{\test}^\axion \rightarrow Q_{\test}^\axion \times m_u m_d/(m_u+m_d)^2$. The numerical values of the relevant axion charges used to impose the MICROSCOPE constraint are
\begin{align}
Q_{\oplus}^\axion &\simeq - 2 \times 10^{-2}, 
\\
|Q_{\text{Pt}}^\axion-Q_{\text{Ti}}^\axion| &\simeq 6 \times 10^{-4}.
\end{align}

When $|\Delta m_{\oplus}| \, R_\oplus < 1$, the axion field only explores the vicinity of the $\phi=0$ vacuum. The axion profile outside the Earth can be obtained by solving the Klein-Gordon equation~\cite{Hees:2018fpg, Banerjee:2022sqg, Gan:2025nlu},
\begin{align}
\label{eq:theta_source_zero_vacuum}
\theta \simeq \theta_0 \cos(m_\phi t) \left( 1 - \frac{ \Delta m_\oplus^2 \, R_\oplus^3}{3 \, r} \right).
\end{align}
Here, $\theta_0 = \phi_0 /f_\phi$ and $\phi_0 = \sqrt{2 \bar{\rho}}/m_\phi$. Substituting Eq.~\eqref{eq:theta_source_zero_vacuum} into Eq.~\eqref{force_axion_general} and taking the time average, we have the attractive force
\begin{equation}
\mathbf{F}_\test \simeq - \frac{\phi_0^2}{8 \pi r^2} \,\frac{(Q_{\test}^\axion M_\test)(Q_{\oplus}^\axion M_\oplus)}{\fphi^4} \rhat,
\end{equation}
which is induced by the background of axion dark matter. Here $M_\test$ and $M_\oplus$ are the masses of the test body and the Earth, respectively. The above formula can also be derived from the quantum field theory~(QFT) as shown in Refs.~\cite{VanTilburg:2024xib,Barbosa:2024pkl,Grossman:2025cov,Cheng:2025fak} under the Born approximation. Since such background-induced force is much weaker than gravity, the corresponding E\"otv\"os parameter is
\bea
\label{eq:eta_axion_appx_pert}
\eta \simeq \frac{\phi_0^2}{8\pi} \frac{|(Q_{\text{Pt}}^\axion - Q_{\text{Ti}}^\axion) \, Q_{\oplus}^\axion| }{ f_\phi^4 \, G_N },
\eea
which depends on the axion density and decay constant. Combining this expression with the experimental constraint on $\eta$, we obtain an upper bound in the $m_\phi$–$f_\phi^{-1}$ plane for axion dark matter. This relation explains the power-law scaling $\fphi^{-1} \propto m_\phi^{1/2}$ in the non-phase-transition regime.

When $|\Delta m_{\oplus}| > R_\oplus^{-1}$ and $|\Delta m_{\oplus}| > m_\phi$, an axion phase transition occurs inside the Earth. In this case, a static axion profile develops around the Earth as shown in Eq.~(\ref{eq:axion_static_profile}). The force acting on each test mass can then be obtained by substituting Eq.~\eqref{eq:axion_static_profile} into Eq.~\eqref{force_axion_general}:
\begin{equation}
\label{eq:Fi_axion_pi_force}
\mathbf{F}_\test \simeq - Q_{\test}^\axion \, M_\test \frac{\pi^2 R_\oplus^2}{r^3} (1+m_\phi r) e^{-2 m_\phi (r-R)} \, \rhat.
\end{equation}
Note that the phase transition takes place only within the Earth and not within the test masses, as $|\Delta m_\test| R_\test \ll 1$, where $R_\test \sim \mathcal{O}(10) \, \text{cm}$ denotes the characteristic size of the test masses. The corresponding E\"otv\"os parameter is
\begin{equation}
\label{eq:eta_axion_appx_pi}
\eta \simeq |Q_{\text{Pt}}^\axion-Q_{\text{Ti}}^\axion| \frac{ \pi^2 R_\oplus^2}{G_N M_\oplus r} \, (1+m_\phi r) \, e^{-2 m_\phi (r-R)},
\end{equation}
where the dependence on $\fphi$ cancels. Since the test-mass accelerations are induced by the static axion field around the Earth, the axion density does not appear explicitly. In the regime $m_\phi \, (r-R_\oplus) \lesssim 1$, we have $\eta \gg 1$, because $R_\oplus/G_N M_\oplus \gg 1$. This implies that, within the Earth's phase-transition regime, the equivalence principle is strongly violated when the Compton wavelength $1/m_\phi$ exceeds the altitude of the MICROSCOPE satellite, $r-R_\oplus$. For $m_\phi \gtrsim 10^{-11}\,\text{eV}$, the force $\mathbf{F}_\test$ becomes exponentially suppressed, as shown in Eq.~(\ref{eq:Fi_axion_pi_force}). Consequently, the E\"otv\"os parameter in Eq.~(\ref{eq:eta_axion_appx_pi}) falls below the experimental limit of Ref.~\cite{MICROSCOPE:2022doy}, and the constraint effectively terminates at $m_\phi \sim 10^{-11}\,\text{eV}$.

To summarize, we show the updated MICROSCOPE constraint on the light QCD axion in Fig.~\ref{fig:main_axion}. For comparison, we also show in Fig.~\ref{fig:MICROSCOPE_QCDAxion} the corresponding constraints for the dilaton with repulsive~\cite{Hees:2018fpg, Banerjee:2022sqg,Gan:2025nlu} and attractive~\cite{Gue:2025nxq,Delaunay:2025pho} potentials. The MICROSCOPE constraint for a pure dilaton is recast into the $m_\phi-f_\phi^{-1}$ plane by neglecting the axion self-interaction. In addition, we include the MICROSCOPE constraint estimated using the QFT approach~\cite{Grossman:2025cov}. From Fig.~\ref{fig:main_axion}, we find that our updated MICROSCOPE constraint agrees with the repulsive and attractive dilaton results in the non-phase-transition regime, where the vacuum is $\phi = 0$, but deviates in the regime where the Earth undergoes a phase transition and a static axion field develops. We also find that our updated constraint differs from that obtained by the QFT approach~\cite{Grossman:2025cov}. The reason is that the latter method does not account for the axion charge difference of two test masses, $\text{Pt}$ and $\text{Ti}$, and the Born approximation becomes invalid when $|\Delta m_\oplus| R_\oplus > 1$.

Before ending the discussion, we briefly comment on other limits when the phase transition takes place inside Earth and a large static axion profile is developed. As noted in Ref.~\cite{Hook:2017psm}, the transition leads to a substantial change of the measured spectrum of hadrons and mesons~\cite{Ubaldi:2008nf}. In addition, as the axion field value approaches close to $\theta \sim \pi$, it will introduce a significant neutron electric dipole moment. This region of parameter space, shaded in blue in Fig.~\ref{fig:MICROSCOPE_QCDAxion}, is likely to be inconsistent with such measurements. MICROSCOPE limit we discuss in this appendix, especially above $m_\phi = 10^{-16}\eV$, provides yet another independent way to probe the same parameter space.

\bibliographystyle{utphys}
\bibliography{ref}
\end{document}